\documentclass[%
reprint,
 amsmath,amssymb,
 aps,
]{revtex4-2}

\usepackage{graphicx}
\usepackage{dcolumn}
\usepackage{bm}
\usepackage{mathtools}

\newcommand{\dd}{\mathrm{d}}

\preprint{APS/123-QED}

\begin{document}

\title{Dynamic structured illumination for confocal microscopy}

\author{Guillaume N\oe{}tinger}
\author{Fabrice Lemoult}%
\author{Sébastien M. Popoff}%
 \email{sebastien.popoff@espci.fr}
\affiliation{Institut Langevin, ESPCI Paris, Université PSL, CNRS, Paris 75005, France}%

\date{\today}

\begin{abstract}
Structured illumination enables the tailoring of an imaging device's optical transfer function to enhance resolution. We propose the incorporation of a temporal periodic modulation, specifically a rotating mask, to encode multiple transfer functions in the temporal domain. This approach is demonstrated using a confocal microscope configuration. At each scanning position, a temporal periodic signal is recorded. By filtering around each harmonic of the rotation frequency, multiple images of the same object can be constructed.
The image carried by the $n{\mathrm{th}}$ harmonic is a convolution of the object with a phase vortex of topological charge $n$, similar to the outcome when using a vortex phase plate as an illumination. This enables the  collection of chosen high spatial frequencies from the sample, thereby enhancing the spatial resolution of the confocal microscope.
\end{abstract}

\maketitle

The optical confocal microscope~\cite{PatentMinsky}, an imaging device extensively utilized for decades, has proven invaluable for scientists investigating phenomena at the scale of hundreds of nanometers. These researchers, including biologists and material scientists, benefit from the device's ability to filter out-of-focus light. This is known as \textit{optical sectioning}. This feature enables the capture of high contrast images even in diffusive samples such as biological tissue~\cite{CorleKino}.  The high-resolution capabilities of the confocal microscope are particularly beneficial for fluorescent imaging~\cite{jonkman_tutorial_2020}. Combined with a depletion beam in the STED 
configuration the device can achieve superresolution, yielding precise structural insights at the cellular level~\cite{STED}. However, fluorescent markers present 
limitations, including their potential toxicity and the prerequisite treatment of the sample, making them unsuitable in some contexts. Consequently, the development 
of optical label-free superresolution microscopy would be highly advantageous in numerous practical applications~\cite{FreeImaging, LabelFreeSuperBook}.

Broadly, the inclusion of time in an optical scheme opens new possibilities~\cite{4DopticsEngheta}.
Mechanical scanning as used in STED, 
illumination and acquisition sequences as seen in STORM~\cite{STORM}, and structured illumination~\cite{StructuredIll}, as well as the analysis of emission fluctuations in SOFI~\cite{SOFI}, all exemplify the prevalent use of time as an additional degree of freedom in numerous superresolution techniques. 
For example, recent work involving illumination modulation in a fluorescent sample with time-varying structured illumination has demonstrated remarkable precision in localization~\cite{ModLoc}.

In this article, we address the challenge of label-free superresolution in the far-field utilizing an
analogous approach that capitalizes on the temporal domain to enhance the volume of data gathered from the object for image reconstruction. To that end, we suggest incorporating wavefront shaping techniques into a standard confocal microscope to introduce a temporal modulation in the signal acquired at each scanning point. The resultant additional degrees of freedom could 
enhance the space-bandwidth product~\cite{SBproductLohmann} of the confocal microscope, leading to an improved resolution.

\section*{Concept}

In the absence of fluorescent probes, the confocal microscope demonstrates a modest improvement in lateral resolution compared to the full-field configuration. In a full-field microscope, the coherent point-spread function (PSF) corresponds to the 2D Fourier transform of the pupil function. With a circular pupil, it is recognized as the Airy function. Owing to the scanning process and under the approximation of a point-like detector, the coherent PSF of the confocal microscope can be expressed as the product of the illumination and collection PSFs~\cite{JMertz}. In a symmetric configuration, its  width is smaller than that of the full field microscope. Using a Gaussian approximation of the PSF, the improvement in lateral resolution is estimated to be on the order of $\sqrt{2}$ which is about $40\%$.

In terms of spatial frequencies, the coherent transfer function (CTF) of a full-field microscope is dictated by the shape of the microscope objective's pupil. For a circular pupil, the CTF takes the form of a circular step function: spatial frequencies with a modulus greater than $N\!A/\lambda$ are filtered out, where $N\!A$ represents the numerical aperture and $\lambda$ is the working wavelength. All spatial frequencies within this zone are transmitted with the same amplitude.
In the confocal microscope, the CTF is the convolution of the illumination and collection CTFs~\cite{SheppardWilson}.
Assuming a symmetric configuration for illumination and collection, the confocal possesses a well-known conical CTF of support twice as large as the full field CTF~\cite{JMertz}. This means that the highest transmitted spatial frequency, $2N\!A/\lambda$, is twice that in the full-field configuration. However, the gain diminishes linearly, peaking at low spatial frequencies and reaching a minimum at the cut-off frequency. In the presence of noise, the low signal to noise ratio at high frequencies leads to a degraded resolution in practice. One straightforward strategy to offset this declining gain is to employ an annular pupil~\cite{AnnularLenseSheppard}. Its drawbacks are a deterioration of the optical sectioning and the presence of secondary lobes. In this paper, we aim to present an alternative approach using a temporal modulation.

\begin{figure}[t!]
\centering\includegraphics[width=1.02\columnwidth]{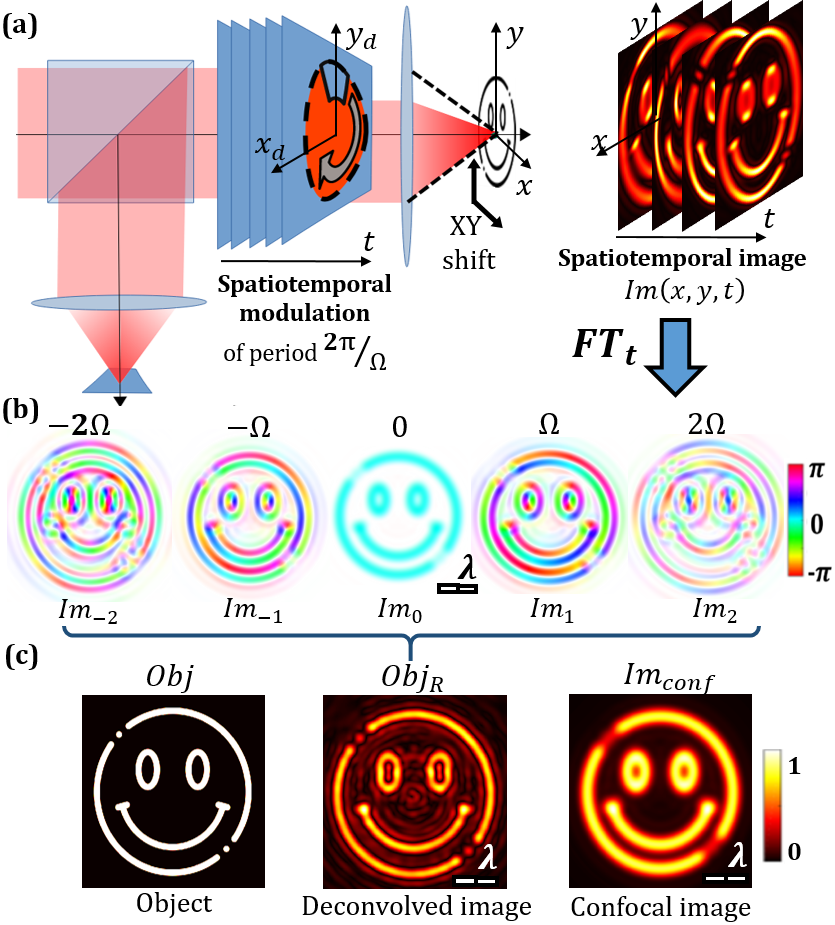}
\caption{\label{fig:SchemeTot} \textbf{Schematic view of the experimental process with simulated data.} \textbf{(a) Acquisition.} A spatiotemporal image is acquired by scanning and temporally modulating the objective's pupil with a rotating mask. \textbf{(b) Analysis.} Thanks to a Fourier series decomposition, the information is gathered in \textit{harmonic} complex images $Im_n$. Each one is carried by a different frequency $n\Omega$ with $n \in \mathbb{Z}$.  \textbf{(c) Reconstruction.} Each image $Im_n$ is deconvolved using a computed PSF. A reconstruction of the sample $Obj_R$ exhibiting sharper details than the confocal image is obtained.} 
\end{figure}

In a recent work, we demonstrated experimentally with acoustic waves that the use of spatiotemporal wavefront shaping allows multiplexing the acquisition for image reconstruction~\cite{PRAppliedAcoustic_Noetinger}. Using a rotating source for the illumination and a rotating receiver for the collection, we obtained different images corresponding to the convolutions of the object with different orthogonal PSFs. The PSFs present different topological phase structures. 
Owing to a periodic Doppler effect, a point-like object perceives a monochromatic wavefield at $\omega_0$ only on the rotation axis and a periodically modulated signal elsewhere equivalent to a frequency comb patterned spectrum. For each frequency $\omega_0+n\Omega$, the field forms a vortex with a vorticity $n$, centered on the optical axis, as though a vortex plate were positioned in the pupil plane. During the backscattering process, the same phenomenon applies, resulting in a focal spot twice as small as that of the full-field microscope at $\omega_0$, and also vortex patterns twice as small as those in the focal plane at other frequencies. This suggests that the rotating emitter and recorder function as a spatiotemporal filter, retaining only the information associated with the high spatial frequency content collected by the confocal microscope. 
Consequently, the presence of harmonic frequencies in the modulated signal perceived by the object, along with vortex-like features, depends solely on the presence of a rotating modulation. Importantly, this is independent of the speed of the modulation relative to the wave's speed or frequency.
By exploiting the diversity of information by summing the images recovered from each PSF, an improvement of the confocal resolution by 70\% is obtained. This improvement enabled the distinction between two point-like objects closer than $\frac{\lambda}{4N\!A}$, surpassing the confocal limit. While the experiment described was implemented using acoustics as a proof of concept, the principles highlighted are broadly applicable to wave-based imaging~\cite{AcousticMicroscope}. In this article, we expand on this approach, applying it to an optical confocal scanning microscope. Nevertheless, it’s crucial to acknowledge that optics possess subtle differences with acoustics that drastically modifies the implementation (see SI). Interestingly, this effect has already been studied in optics for the detection of rotating bodies in astronomy \cite{LaveryRotDoppler} but to our knowledge has never been applied in microscopy. 


To engineer a high-speed, time-varying illumination with optical waves that will not significantly impede the confocal acquisition process, we opt to utilize a Digital Micromirror Device (DMD) optically conjugated to the pupil plane of a microscope objective (Figure~\ref{fig:SchemeTot}\textbf{.a}). This device enables amplitude modulation of the field at approximately 10~kHz. The time-varying pattern displayed
modulates temporally the pupil function and, as a result, the CTF. In this scenario, both the illumination and collection pupils are time-varying, which is equivalent to a rotation of the object via a change of frame.  
In the following, we focus on a particular type of illumination consisting on a pattern rotating about the optical axis.
We demonstrate how it allows multiplexing the image acquisition process, enabling the efficient extraction of the highest spatial frequency components. We show the capacity of this approach to improve the image resolution.

\section*{Numerical approach}

\begin{figure}[t!]
\centering\includegraphics[width=1.02\columnwidth]{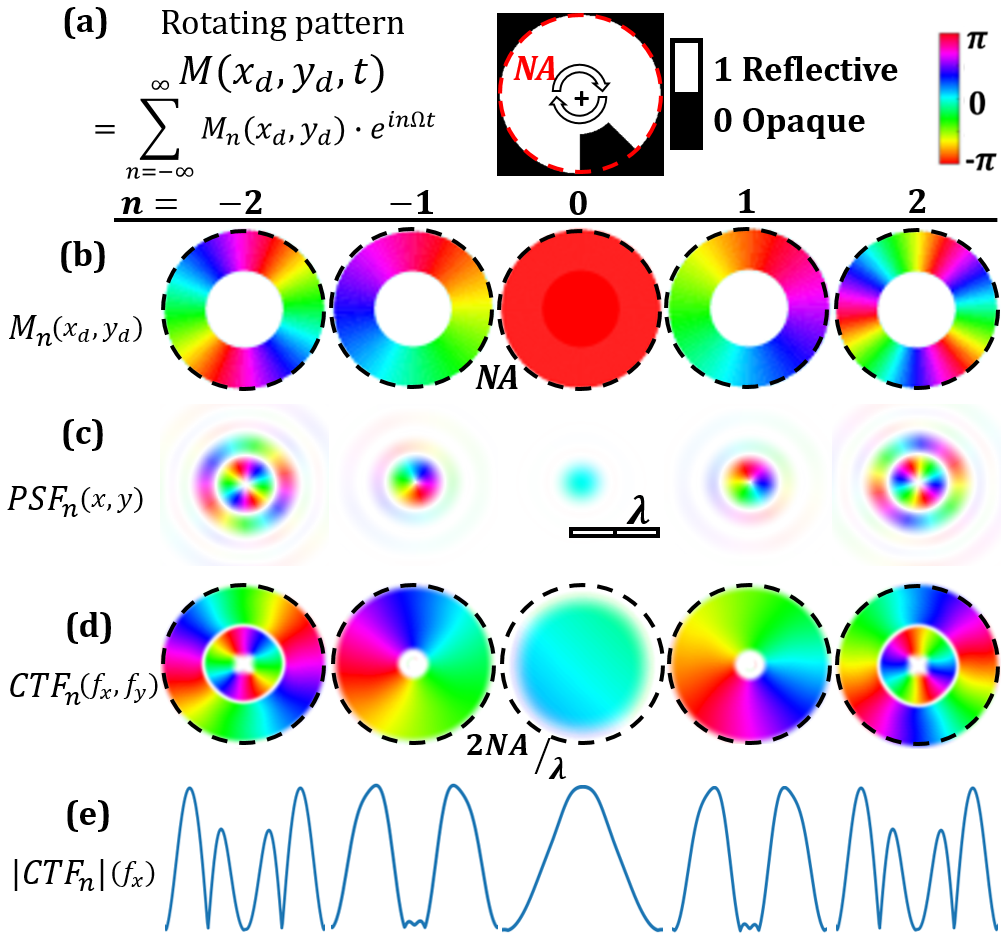}
\caption{\label{fig:PSF_OTF} \textbf{PSF and CTF of the system associated to a given pattern} decomposed as a Fourier series on the frequencies $n\Omega$ with $n\in [\![-2,2]\!]$. \textbf{(a)} The rotating pattern $M$ sent to the DMD. \textbf{(b) Decomposition of $M(x_d,y_d,t)$}. The pupil is a linear combination of vortex plates. \textbf{(c) PSF} in the sample's plane. \textbf{(d) Coherent transfer functions} ($CTF_n$) in the spatial frequency space. The dotted line represents the confocal resolution limit $2\lambda/N\!A$ \textbf{(e) Cross section view of $CTF_n$} modulus. The low spatial frequencies are not collected for values of $n$ different than 0, $CTF_0$ corresponds roughly to the classical confocal CTF.}
\end{figure}

We choose a sequence on the DMD, $M(x_d,y_d,t)$, consisting of the full pupil deprived from a 45$^\circ$ truncated sector as depicted on Figure \ref{fig:PSF_OTF}.\textbf{a}. This preserves the optical sectioning as well as providing some signal amplification of the temporal backscattered signal associated to the rotating pattern (see SI).

As a time-periodic pattern, it can be decomposed as a Fourier series with coefficients $M_n(x_d,y_d)$. As seen in Figure \ref{fig:PSF_OTF}\textbf{.b}, these coefficients are associated to vortices. Similarly, the corresponding temporal PSF or CTF can be computed and then again expressed as Fourier series (Figure~\ref{fig:PSF_OTF}\textbf{.c} \& \textbf{d}). As a topological invariant~\cite{VortexProp}, the vorticity seen on the DMD patterns is conserved and is seen on the PSFs; thus guaranteeing their orthogonality. Indeed, the PSF corresponding to the frequency $n\Omega$ is a vortex of vorticity $n$ with a radius increasing with $|n|$. $CTF_0$ is roughly equivalent to the confocal CTF: the average pupil used during the illumination being almost the full microscope objective pupil, $CTF_0$ is also roughly the autoconvolution of the objective's pupil. The other dynamic CTFs possess a vorticity which imposes a zero for the low spatial frequencies.  
$CTF_{n \neq 0}$ carries information with a high gain only for the high spatial frequencies of the sample.
This illustrates the benefit of using a rotating illumination in confocal microscopy since those frequencies are usually transmitted with a low gain leading to a resolution lower than $\frac{2N\!A}{\lambda}$ in the presence of noise. 

\section*{Experiment \& Results}

Let us examine a practical implementation of the experiment (Figure~\ref{fig:SchemeTot}) detailed in SI. It is made using a narrowband polarized laser Coherent Sapphire SF NX @488~nm. The beam is enlarged with beam expanders, filtered using a pinhole, sent to a 2~Mpx~\textit{Vialux} DMD and then to the sample using an \textit{Olympus} MPLFLN40X microscope objective of numerical aperture $N\!A=0.75$. With a quarter waveplate, the flux is sent to a \textit{Thorlabs} PDA10A2 photodiode after a second passage by the polarizing beam splitter cube. The latter is placed behind a pinhole whose equivalent size in the object plane is approximately 1~Airy unit. The current from the photodiode is amplified by a transimpedance amplifier and recorded by a \textit{Picoscope} electronic oscilloscope.

Using a 1951 \textit{USAF} target, the full-field resolution in white light is near 388~nm, close to $\lambda/2N\!A=325$~nm and the confocal resolution is determined to be 244~nm close enough to $\frac{\lambda}{2\sqrt{2}}=230~nm$ to consider the set-up to be diffraction-limited (see SI).

\begin{figure}[b!]
\centering\includegraphics[width=1.02\columnwidth]{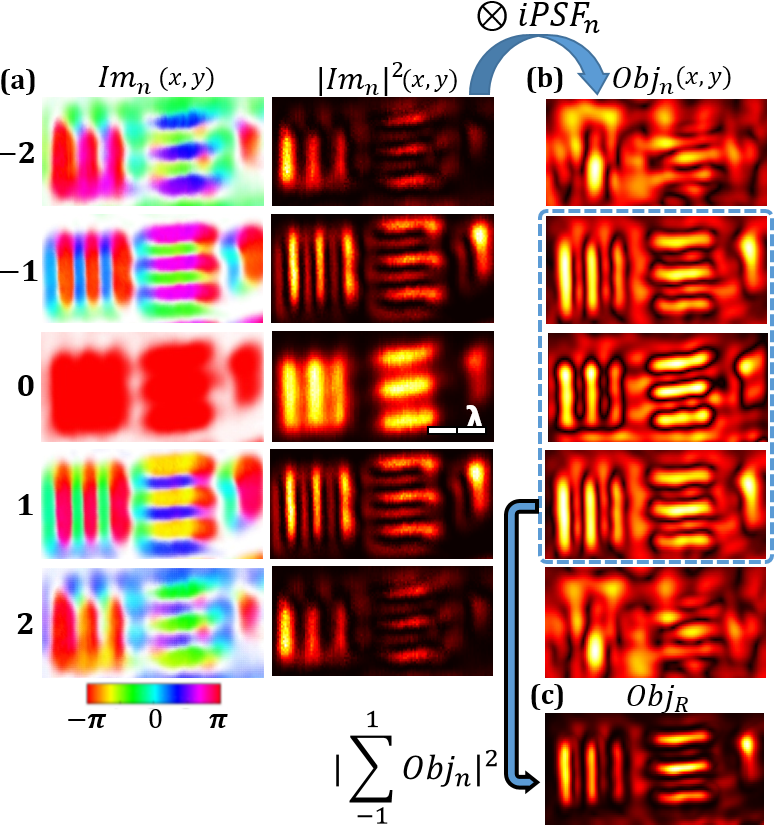}
\caption{\label{fig:ImDynG11e1}  \textbf{Illustration of the deconvolution procedure with experimental images} of a ReadyOptics USAF resolution target group 11 element 1 (244nm). 
\textbf{(a)}~Dynamic confocal images obtained with the pattern of Figure \ref{fig:PSF_OTF}\textbf{.a} at different harmonic frequencies $n \Omega$ for $n \in [\![-2;2]\!]$ with weighted phase and intensity. \textbf{(b)}~Deconvolved confocal dynamic images with numerical pseudoinverse $iPSF_n$. \textbf{(c)}~Sum of $Obj_n$ with $n \in [\![-1;1]\!]$ to obtain a final reconstruction $Obj_R$.}
\end{figure}

For each scan point, the sequence of 60 masks $M(x_d,y_d,t)$ is displayed first followed by a circular pattern associated to the full aperture. In this way, dynamic and \textit{standard} confocal images are acquired with the same scan.
We employ the sequence depicted in Figure~\ref{fig:PSF_OTF}\textbf{.a} 
to capture images of parallel lines from the USAF resolution test chart, 
each line featuring a thickness and separation distance of 244~nm. After temporal Fourier decomposition of the received signals on the photodiode we are capable to retrieve images at the different harmonics $n\Omega$ with $n \in [\![-2;2]\!]$ (Figure~\ref{fig:ImDynG11e1}.\textbf{a)}). Each image provides a phase information that is reminiscent of the vortex nature of each PSF, except for the image at $n=0$ which is equivalent to an intensity image. Each of these images carry different informations of the same objects with its own noise.

For absolute values of $n$ exceeding 1, experimental data begin to diverge from the theoretical and numerical predictions. A possibility is that higher order vortices break into $\pm 1$ vortices that are the only one existing naturally~\cite{RoadmapStructuredLightVortex} when encountering the sample. Other possible explanations are the integrating effect of the pinhole~\cite{SheppardWilson} which is in fact of finite size and the setup's susceptibility to misalignment, thermal instability, and mechanical vibrations as reported in a similar experiment~\cite{SuperresLinearPushkina}.\\

To build a single image out of this series, we implement a simple inversion procedure. Each image at each harmonic is deconvolved by its own inverted PSF predicted by the numerical simulation. To ensure the procedure is resistant to noise, a Tikhonov regularization is employed during the inversion~\cite{TikhReg,PopoffOpaqueTikh}. The regularized inverse operator writes:
\begin{equation}
    iCTF_n=i\widehat{PSF}_n=(CTF_n^* \cdot CTF_n + \sigma)^{-1}\cdot CTF_n^* \label{InvTikh}
\end{equation}
\noindent $\sigma$ being the noise-to-signal ratio, $\cdot^*$ denotes the complex conjugate and $\widehat{{\:.\:}}$ the 2D spatial Fourier transform. Note that the pseudo-inverse of a vortex-like PSF is also a vortex-like function with the opposite topological charge. Each image from each harmonic yields a deconvolved image referred to as $Obj_n$, 
as depicted in Figure \ref{fig:ImDynG11e1}\textbf{.b)}. 
Each image resembles the target object, 
albeit with noticeable degradation for $n=2$, 
aligning with the discrepancies observed earlier.
The final image is achieved by summing all the deconvolved intensity images, 
resulting in a pattern of enhanced contrast and improved resolution.

\begin{figure}[t!]
    \begin{center}
        \includegraphics[width=0.85\columnwidth]{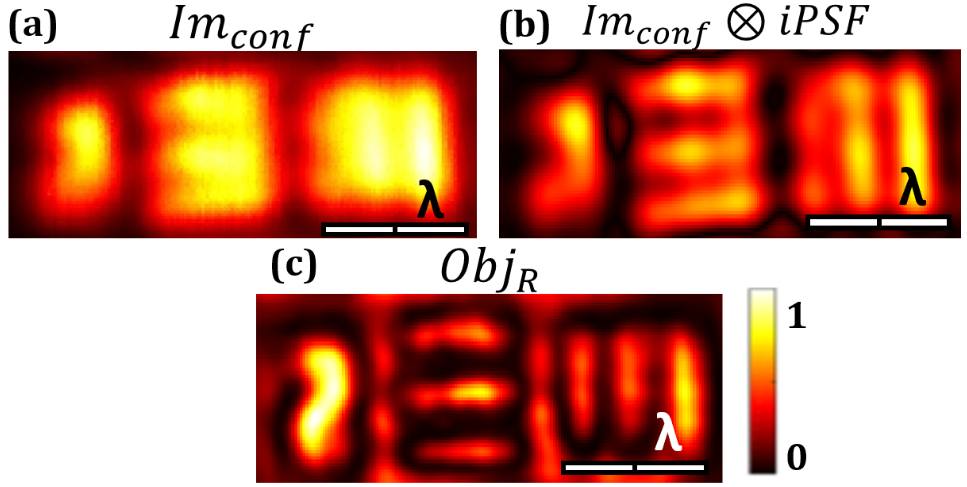}
        \caption{\label{fig:ImGr11e2}\textbf{Resolution limit}. \textbf{a)} Regular confocal image of group 11 element 2 of USAF target, linewidth is 218nm. \textbf{b)} Deconvolution of the confocal image. \textbf{c)} Reconstruction obtained by summing the deconvolved dynamic confocal images obtained with our method. This is a 10\% improvement in resolution.} 
    \end{center}
\end{figure}

To draw a comparison with the standard confocal setup, 
we present in Figure~\ref{fig:ImGr11e2} the confocal images, 
both with and without inversion, 
alongside the results from our approach, 
pertaining to lines that are now 218~nm thick, a bit smaller than in the previous Figure. 
The reconstructed image resulting from the temporally modulated wavefronts is the only one that successfully allows discriminating the individual lines. This represents a 10\% improvement in resolution compared to the inverted confocal image. 


\section*{Conclusion \& Perspectives}

In this article, we provide a proof-of-concept of adding a temporal modulation to an imaging scheme. As an example, we chose to use rotating wavefronts to enhance the lateral resolution of confocal microscopy. Displaying a pattern rotating by the optical axis 
leads to the same periodic modulation of the illumination  equivalent to a frequency comb observed with sound waves. For each frequency, a different image of the same sample is obtained. All the information can be summed to obtain an improved reconstruction of the object. Our results demonstrate this technique's capacity to improve the contrast and resolution of confocal imaging. 

Although our experiments suffer from the current flaws of the wavefront shaping tools as highlighted in~\cite{RoadmapStructuredLight}, namely, the slowness of SLMs and the binary amplitude modulation of DMDs (not to mention the introduction of mechanical vibrations and aberrations), our approach opens new perspectives for confocal imaging. 
While it has already outperformed confocal imaging in terms of resolution, 
we anticipate that addressing stability concerns and 
implementing a more sophisticated reconstruction algorithm could lead to 
further enhancements in resolution and image quality. 
In the study presented, we leverage the vortex shape at each harmonic,
which facilitates the selective reconstruction of information 
from different regions of the image's spatial spectrum to enhance the image quality.
Noticeably, the implementation of spatiotemporal modulation 
unlocks these new degrees of freedom, 
while still necessitating only a single photodetector for the measurement part.
This application is an example of PSF engineering 
as we exploit the spatio-temporal modulation of the illumination field 
to generate controlled CTFs for different harmonics.
It offers control over the reconstruction process, 
enabling one to concentrate,
for example, on parts of the spectrum more sensitive to noise, 
or to selectively detect specific patterns.

Moreover, these ideas could be applied to various optical setups in a full-field configuration, \textit{e.g.} readily in image scanning microscopy \cite{ImageScanningMicroscopy} to enhance again the resolution. In astronomy, a possible application could be to replace the vortex plates used in coronagraphy~\cite{CoronaLightValve} with more flexibility.\\

\small{ \paragraph*{Funding}
This work has received support under the program ''\textit{Investissements d’Avenir}'' launched by the French Government, and is partially supported by the Simons Foundation/Collaboration on Symmetry-Driven Extreme Wave Phenomena.\\
\paragraph*{Data availability} Data underlying the results presented here may be obtained from the authors upon reasonable request.}



\bibliography{apssamp}

\begin{thebibliography}{38}%
\makeatletter
\providecommand \@ifxundefined [1]{%
 \@ifx{#1\undefined}
}%
\providecommand \@ifnum [1]{%
 \ifnum #1\expandafter \@firstoftwo
 \else \expandafter \@secondoftwo
 \fi
}%
\providecommand \@ifx [1]{%
 \ifx #1\expandafter \@firstoftwo
 \else \expandafter \@secondoftwo
 \fi
}%
\providecommand \natexlab [1]{#1}%
\providecommand \enquote  [1]{``#1''}%
\providecommand \bibnamefont  [1]{#1}%
\providecommand \bibfnamefont [1]{#1}%
\providecommand \citenamefont [1]{#1}%
\providecommand \href@noop [0]{\@secondoftwo}%
\providecommand \href [0]{\begingroup \@sanitize@url \@href}%
\providecommand \@href[1]{\@@startlink{#1}\@@href}%
\providecommand \@@href[1]{\endgroup#1\@@endlink}%
\providecommand \@sanitize@url [0]{\catcode `\\12\catcode `\$12\catcode
  `\&12\catcode `\#12\catcode `\^12\catcode `\_12\catcode `\%12\relax}%
\providecommand \@@startlink[1]{}%
\providecommand \@@endlink[0]{}%
\providecommand \url  [0]{\begingroup\@sanitize@url \@url }%
\providecommand \@url [1]{\endgroup\@href {#1}{\urlprefix }}%
\providecommand \urlprefix  [0]{URL }%
\providecommand \Eprint [0]{\href }%
\providecommand \doibase [0]{https://doi.org/}%
\providecommand \selectlanguage [0]{\@gobble}%
\providecommand \bibinfo  [0]{\@secondoftwo}%
\providecommand \bibfield  [0]{\@secondoftwo}%
\providecommand \translation [1]{[#1]}%
\providecommand \BibitemOpen [0]{}%
\providecommand \bibitemStop [0]{}%
\providecommand \bibitemNoStop [0]{.\EOS\space}%
\providecommand \EOS [0]{\spacefactor3000\relax}%
\providecommand \BibitemShut  [1]{\csname bibitem#1\endcsname}%
\let\auto@bib@innerbib\@empty
\bibitem [{\citenamefont {{Marvin Minsky}}(1957)}]{PatentMinsky}%
  \BibitemOpen
  \bibfield  {author} {\bibinfo {author} {\bibnamefont {{Marvin Minsky}}},\
  }\href@noop {} {\bibinfo {title} {{Microscopy apparatus}}} (\bibinfo {year}
  {{1957}})\BibitemShut {NoStop}%
\bibitem [{\citenamefont {Corle}\ and\ \citenamefont {Kino}(1996)}]{CorleKino}%
  \BibitemOpen
  \bibfield  {author} {\bibinfo {author} {\bibfnamefont {T.~R.}\ \bibnamefont
  {Corle}}\ and\ \bibinfo {author} {\bibfnamefont {G.}~\bibnamefont {Kino}},\
  }\href@noop {} {\emph {\bibinfo {title} {Confocal Scanning Optical Microscopy
  and Related Imaging Systems}}}\ (\bibinfo  {publisher} {Academic Press},\
  \bibinfo {year} {1996})\BibitemShut {NoStop}%
\bibitem [{\citenamefont {Jonkman}\ \emph {et~al.}(2020)\citenamefont
  {Jonkman}, \citenamefont {Brown}, \citenamefont {Wright}, \citenamefont
  {Anderson},\ and\ \citenamefont {North}}]{jonkman_tutorial_2020}%
  \BibitemOpen
  \bibfield  {author} {\bibinfo {author} {\bibfnamefont {J.}~\bibnamefont
  {Jonkman}}, \bibinfo {author} {\bibfnamefont {C.~M.}\ \bibnamefont {Brown}},
  \bibinfo {author} {\bibfnamefont {G.~D.}\ \bibnamefont {Wright}}, \bibinfo
  {author} {\bibfnamefont {K.~I.}\ \bibnamefont {Anderson}},\ and\ \bibinfo
  {author} {\bibfnamefont {A.~J.}\ \bibnamefont {North}},\ }\bibfield  {title}
  {\bibinfo {title} {Tutorial: guidance for quantitative confocal microscopy},\
  }\href {https://doi.org/10.1038/s41596-020-0313-9} {\bibfield  {journal}
  {\bibinfo  {journal} {Nature Protocols}\ }\textbf {\bibinfo {volume} {15}},\
  \bibinfo {pages} {1585} (\bibinfo {year} {2020})}\BibitemShut {NoStop}%
\bibitem [{\citenamefont {Klar}\ \emph {et~al.}(2006)\citenamefont {Klar},
  \citenamefont {Engel},\ and\ \citenamefont {Hell}}]{STED}%
  \BibitemOpen
  \bibfield  {author} {\bibinfo {author} {\bibfnamefont {T.}~\bibnamefont
  {Klar}}, \bibinfo {author} {\bibfnamefont {E.}~\bibnamefont {Engel}},\ and\
  \bibinfo {author} {\bibfnamefont {S.}~\bibnamefont {Hell}},\ }\bibfield
  {title} {\bibinfo {title} {Breaking abbes diffraction resolution limit in
  fluorescence microscopy with stimulated emission depletion beams of various
  shapes.},\ }\href@noop {} {\bibfield  {journal} {\bibinfo  {journal} {Phys.
  Rev. E}\ }\textbf {\bibinfo {volume} {24}},\ \bibinfo {pages} {793} (\bibinfo
  {year} {2006})}\BibitemShut {NoStop}%
\bibitem [{\citenamefont {Marx}(2019)}]{FreeImaging}%
  \BibitemOpen
  \bibfield  {author} {\bibinfo {author} {\bibfnamefont {V.}~\bibnamefont
  {Marx}},\ }\bibfield  {title} {\bibinfo {title} {It’s free imaging —
  label-free, that is.},\ }\href@noop {} {\bibfield  {journal} {\bibinfo
  {journal} {Nat Methods}\ }\textbf {\bibinfo {volume} {16}},\ \bibinfo {pages}
  {1209} (\bibinfo {year} {2019})}\BibitemShut {NoStop}%
\bibitem [{\citenamefont {Astratov}(2019)}]{LabelFreeSuperBook}%
  \BibitemOpen
  \bibfield  {author} {\bibinfo {author} {\bibfnamefont {V.}~\bibnamefont
  {Astratov}},\ }\href@noop {} {\emph {\bibinfo {title} {Label-Free
  Super-Resolution Microscopy}}}\ (\bibinfo  {publisher} {Springer},\ \bibinfo
  {year} {2019})\BibitemShut {NoStop}%
\bibitem [{\citenamefont {Engheta}(2023)}]{4DopticsEngheta}%
  \BibitemOpen
  \bibfield  {author} {\bibinfo {author} {\bibfnamefont {N.}~\bibnamefont
  {Engheta}},\ }\bibfield  {title} {\bibinfo {title} {Four-dimensional optics
  using time-varying metamaterials},\ }\href
  {https://doi.org/10.1126/science.adf1094} {\bibfield  {journal} {\bibinfo
  {journal} {Science}\ }\textbf {\bibinfo {volume} {379}},\ \bibinfo {pages}
  {1190} (\bibinfo {year} {2023})}\BibitemShut {NoStop}%
\bibitem [{\citenamefont {Rust}\ \emph {et~al.}(2001)\citenamefont {Rust},
  \citenamefont {Bates},\ and\ \citenamefont {Zhuang.}}]{STORM}%
  \BibitemOpen
  \bibfield  {author} {\bibinfo {author} {\bibfnamefont {M.}~\bibnamefont
  {Rust}}, \bibinfo {author} {\bibfnamefont {M.}~\bibnamefont {Bates}},\ and\
  \bibinfo {author} {\bibfnamefont {X.}~\bibnamefont {Zhuang.}},\ }\bibfield
  {title} {\bibinfo {title} {Sub-diffraction-limit imaging by stochastic
  optical reconstruction microscopy (storm).},\ }\href@noop {} {\bibfield
  {journal} {\bibinfo  {journal} {Nature Methods}\ }\textbf {\bibinfo {volume}
  {3}} (\bibinfo {year} {2001})}\BibitemShut {NoStop}%
\bibitem [{\citenamefont {Gustafsson}(2001)}]{StructuredIll}%
  \BibitemOpen
  \bibfield  {author} {\bibinfo {author} {\bibfnamefont {M.~G.~L.}\
  \bibnamefont {Gustafsson}},\ }\bibfield  {title} {\bibinfo {title}
  {Surpassing the lateral resolution limit by a factor of two using structured
  illumination microscopy},\ }\href@noop {} {\bibfield  {journal} {\bibinfo
  {journal} {Journal of Microscopy}\ }\textbf {\bibinfo {volume} {198}},\
  \bibinfo {pages} {82} (\bibinfo {year} {2001})}\BibitemShut {NoStop}%
\bibitem [{\citenamefont {Dertinger}\ \emph {et~al.}(2009)\citenamefont
  {Dertinger}, \citenamefont {Colyer}, \citenamefont {Iyer}, \citenamefont
  {Weiss},\ and\ \citenamefont {Enderlein}}]{SOFI}%
  \BibitemOpen
  \bibfield  {author} {\bibinfo {author} {\bibfnamefont {T.}~\bibnamefont
  {Dertinger}}, \bibinfo {author} {\bibfnamefont {R.}~\bibnamefont {Colyer}},
  \bibinfo {author} {\bibfnamefont {G.}~\bibnamefont {Iyer}}, \bibinfo {author}
  {\bibfnamefont {S.}~\bibnamefont {Weiss}},\ and\ \bibinfo {author}
  {\bibfnamefont {J.}~\bibnamefont {Enderlein}},\ }\bibfield  {title} {\bibinfo
  {title} {Fast, background-free, 3d super-resolution optical fluctuation
  imaging (sofi)},\ }\href@noop {} {\bibfield  {journal} {\bibinfo  {journal}
  {PNAS.}\ ,\ \bibinfo {pages} {22287}} (\bibinfo {year} {2009})}\BibitemShut
  {NoStop}%
\bibitem [{\citenamefont {Jouchet}\ \emph {et~al.}(2021)\citenamefont
  {Jouchet}, \citenamefont {Cabriel}, \citenamefont {Bourg}, \citenamefont
  {Bardou}, \citenamefont {Poüs}, \citenamefont {Fort},\ and\ \citenamefont
  {Lévêque-Fort}}]{ModLoc}%
  \BibitemOpen
  \bibfield  {author} {\bibinfo {author} {\bibfnamefont {P.}~\bibnamefont
  {Jouchet}}, \bibinfo {author} {\bibfnamefont {C.}~\bibnamefont {Cabriel}},
  \bibinfo {author} {\bibfnamefont {N.}~\bibnamefont {Bourg}}, \bibinfo
  {author} {\bibfnamefont {M.}~\bibnamefont {Bardou}}, \bibinfo {author}
  {\bibfnamefont {C.}~\bibnamefont {Poüs}}, \bibinfo {author} {\bibfnamefont
  {E.}~\bibnamefont {Fort}},\ and\ \bibinfo {author} {\bibfnamefont
  {S.}~\bibnamefont {Lévêque-Fort}},\ }\bibfield  {title} {\bibinfo {title}
  {Nanometric axial localization of single fluorescent molecules with modulated
  excitation.},\ }\href@noop {} {\bibfield  {journal} {\bibinfo  {journal}
  {Nat. Photonics}\ ,\ \bibinfo {pages} {297}} (\bibinfo {year}
  {2021})}\BibitemShut {NoStop}%
\bibitem [{\citenamefont {Lohmann}\ \emph {et~al.}(1996)\citenamefont
  {Lohmann}, \citenamefont {Dorsch}, \citenamefont {Mendlovic}, \citenamefont
  {Zalevsky},\ and\ \citenamefont {Ferreira}}]{SBproductLohmann}%
  \BibitemOpen
  \bibfield  {author} {\bibinfo {author} {\bibfnamefont {A.~W.}\ \bibnamefont
  {Lohmann}}, \bibinfo {author} {\bibfnamefont {R.~G.}\ \bibnamefont {Dorsch}},
  \bibinfo {author} {\bibfnamefont {D.}~\bibnamefont {Mendlovic}}, \bibinfo
  {author} {\bibfnamefont {Z.}~\bibnamefont {Zalevsky}},\ and\ \bibinfo
  {author} {\bibfnamefont {C.}~\bibnamefont {Ferreira}},\ }\bibfield  {title}
  {\bibinfo {title} {Space--bandwidth product of optical signals and systems},\
  }\href {https://doi.org/10.1364/JOSAA.13.000470} {\bibfield  {journal}
  {\bibinfo  {journal} {J. Opt. Soc. Am. A}\ }\textbf {\bibinfo {volume}
  {13}},\ \bibinfo {pages} {470} (\bibinfo {year} {1996})}\BibitemShut
  {NoStop}%
\bibitem [{\citenamefont {Mertz}(2019)}]{JMertz}%
  \BibitemOpen
  \bibfield  {author} {\bibinfo {author} {\bibfnamefont {J.}~\bibnamefont
  {Mertz}},\ }\href@noop {} {\emph {\bibinfo {title} {Introduction to Optical
  Microscopy}}},\ \bibinfo {edition} {2nd}\ ed.\ (\bibinfo  {publisher}
  {Cambridge University Press},\ \bibinfo {year} {2019})\BibitemShut {NoStop}%
\bibitem [{\citenamefont {Wilson}\ and\ \citenamefont
  {Sheppard}(1984)}]{SheppardWilson}%
  \BibitemOpen
  \bibfield  {author} {\bibinfo {author} {\bibfnamefont {T.}~\bibnamefont
  {Wilson}}\ and\ \bibinfo {author} {\bibfnamefont {C.}~\bibnamefont
  {Sheppard}},\ }\href@noop {} {\emph {\bibinfo {title} {Theory and Practice of
  Scanning Confocal Microscopy}}}\ (\bibinfo  {publisher} {Academic Press},\
  \bibinfo {year} {1984})\BibitemShut {NoStop}%
\bibitem [{\citenamefont {Sheppard}\ and\ \citenamefont
  {Wilson}(1979)}]{AnnularLenseSheppard}%
  \BibitemOpen
  \bibfield  {author} {\bibinfo {author} {\bibfnamefont {C.~J.~R.}\
  \bibnamefont {Sheppard}}\ and\ \bibinfo {author} {\bibfnamefont
  {T.}~\bibnamefont {Wilson}},\ }\bibfield  {title} {\bibinfo {title} {Imaging
  properties of annular lenses},\ }\href {https://doi.org/10.1364/AO.18.003764}
  {\bibfield  {journal} {\bibinfo  {journal} {Appl. Opt.}\ }\textbf {\bibinfo
  {volume} {18}},\ \bibinfo {pages} {3764} (\bibinfo {year}
  {1979})}\BibitemShut {NoStop}%
\bibitem [{\citenamefont {Noetinger}\ \emph {et~al.}(2023)\citenamefont
  {Noetinger}, \citenamefont {M\'etais}, \citenamefont {Lerosey}, \citenamefont
  {Fink}, \citenamefont {Popoff},\ and\ \citenamefont
  {Lemoult}}]{PRAppliedAcoustic_Noetinger}%
  \BibitemOpen
  \bibfield  {author} {\bibinfo {author} {\bibfnamefont {G.}~\bibnamefont
  {Noetinger}}, \bibinfo {author} {\bibfnamefont {S.}~\bibnamefont {M\'etais}},
  \bibinfo {author} {\bibfnamefont {G.}~\bibnamefont {Lerosey}}, \bibinfo
  {author} {\bibfnamefont {M.}~\bibnamefont {Fink}}, \bibinfo {author}
  {\bibfnamefont {S.~M.}\ \bibnamefont {Popoff}},\ and\ \bibinfo {author}
  {\bibfnamefont {F.}~\bibnamefont {Lemoult}},\ }\bibfield  {title} {\bibinfo
  {title} {Superresolved imaging based on spatiotemporal wave-front shaping},\
  }\href {https://doi.org/10.1103/PhysRevApplied.19.024032} {\bibfield
  {journal} {\bibinfo  {journal} {Phys. Rev. Appl.}\ }\textbf {\bibinfo
  {volume} {19}},\ \bibinfo {pages} {024032} (\bibinfo {year}
  {2023})}\BibitemShut {NoStop}%
\bibitem [{\citenamefont {Weglein}\ and\ \citenamefont
  {Wilson}(1977)}]{AcousticMicroscope}%
  \BibitemOpen
  \bibfield  {author} {\bibinfo {author} {\bibfnamefont {R.~D.}\ \bibnamefont
  {Weglein}}\ and\ \bibinfo {author} {\bibfnamefont {R.~G.}\ \bibnamefont
  {Wilson}},\ }\bibfield  {title} {\bibinfo {title} {Image resolution of the
  scanning acoustic microscope.},\ }\href@noop {} {\bibfield  {journal}
  {\bibinfo  {journal} {Appl. Phys. Lett.}\ }\textbf {\bibinfo {volume} {31}},\
  \bibinfo {pages} {793} (\bibinfo {year} {1977})}\BibitemShut {NoStop}%
\bibitem [{\citenamefont {Lavery}\ \emph {et~al.}(2013)\citenamefont {Lavery},
  \citenamefont {Speirits}, \citenamefont {Barnett},\ and\ \citenamefont
  {Padgett}}]{LaveryRotDoppler}%
  \BibitemOpen
  \bibfield  {author} {\bibinfo {author} {\bibfnamefont {M.~P.~J.}\
  \bibnamefont {Lavery}}, \bibinfo {author} {\bibfnamefont {F.~C.}\
  \bibnamefont {Speirits}}, \bibinfo {author} {\bibfnamefont {S.~M.}\
  \bibnamefont {Barnett}},\ and\ \bibinfo {author} {\bibfnamefont {M.~J.}\
  \bibnamefont {Padgett}},\ }\bibfield  {title} {\bibinfo {title} {Detection of
  a spinning object using light's orbital angular momentum},\ }\href
  {https://doi.org/10.1126/science.1239936} {\bibfield  {journal} {\bibinfo
  {journal} {Science}\ }\textbf {\bibinfo {volume} {341}},\ \bibinfo {pages}
  {537} (\bibinfo {year} {2013})}\BibitemShut {NoStop}%
\bibitem [{\citenamefont {G.}(1993)}]{VortexProp}%
  \BibitemOpen
  \bibfield  {author} {\bibinfo {author} {\bibfnamefont {I.}~\bibnamefont
  {G.}},\ }\bibfield  {title} {\bibinfo {title} {Optical vortices and their
  propagation},\ }\href {https://doi.org/10.1080/09500349314550101} {\bibfield
  {journal} {\bibinfo  {journal} {Journal of Modern Optics}\ }\textbf {\bibinfo
  {volume} {40}},\ \bibinfo {pages} {73} (\bibinfo {year} {1993})}\BibitemShut
  {NoStop}%
\bibitem [{\citenamefont {Pushkina}\ \emph {et~al.}(2021)\citenamefont
  {Pushkina}, \citenamefont {Maltese}, \citenamefont {Costa-Filho},
  \citenamefont {Patel},\ and\ \citenamefont
  {Lvovsky}}]{SuperresLinearPushkina}%
  \BibitemOpen
  \bibfield  {author} {\bibinfo {author} {\bibfnamefont {A.~A.}\ \bibnamefont
  {Pushkina}}, \bibinfo {author} {\bibfnamefont {G.}~\bibnamefont {Maltese}},
  \bibinfo {author} {\bibfnamefont {J.~I.}\ \bibnamefont {Costa-Filho}},
  \bibinfo {author} {\bibfnamefont {P.}~\bibnamefont {Patel}},\ and\ \bibinfo
  {author} {\bibfnamefont {A.~I.}\ \bibnamefont {Lvovsky}},\ }\bibfield
  {title} {\bibinfo {title} {Superresolution linear optical imaging in the far
  field},\ }\href {https://doi.org/10.1103/PhysRevLett.127.253602} {\bibfield
  {journal} {\bibinfo  {journal} {Phys. Rev. Lett.}\ }\textbf {\bibinfo
  {volume} {127}},\ \bibinfo {pages} {253602} (\bibinfo {year}
  {2021})}\BibitemShut {NoStop}%
\bibitem [{\citenamefont {Tikhonov}(1963)}]{TikhReg}%
  \BibitemOpen
  \bibfield  {author} {\bibinfo {author} {\bibfnamefont {A.}~\bibnamefont
  {Tikhonov}},\ }\bibfield  {title} {\bibinfo {title} {Solution of incorrectly
  formulated problems and the regularization method},\ }\href@noop {}
  {\bibfield  {journal} {\bibinfo  {journal} {Soviet Math. Dokl.}\ }\textbf
  {\bibinfo {volume} {4}},\ \bibinfo {pages} {1035} (\bibinfo {year}
  {1963})}\BibitemShut {NoStop}%
\bibitem [{\citenamefont {Popoff}\ \emph {et~al.}(2010)\citenamefont {Popoff},
  \citenamefont {Lerosey}, \citenamefont {Fink}, \citenamefont {Boccara},\ and\
  \citenamefont {Gigan}}]{PopoffOpaqueTikh}%
  \BibitemOpen
  \bibfield  {author} {\bibinfo {author} {\bibfnamefont {S.}~\bibnamefont
  {Popoff}}, \bibinfo {author} {\bibfnamefont {G.}~\bibnamefont {Lerosey}},
  \bibinfo {author} {\bibfnamefont {M.}~\bibnamefont {Fink}}, \bibinfo {author}
  {\bibfnamefont {A.~C.}\ \bibnamefont {Boccara}},\ and\ \bibinfo {author}
  {\bibfnamefont {S.}~\bibnamefont {Gigan}},\ }\bibfield  {title} {\bibinfo
  {title} {Image transmission through an opaque material},\ }\href@noop {}
  {\bibfield  {journal} {\bibinfo  {journal} {Nature Comm.}\ }\textbf {\bibinfo
  {volume} {1}},\ \bibinfo {pages} {81} (\bibinfo {year} {2010})}\BibitemShut
  {NoStop}%
\bibitem [{\citenamefont {Ritsch-Marte}(2017)}]{RoadmapStructuredLight}%
  \BibitemOpen
  \bibfield  {author} {\bibinfo {author} {\bibfnamefont {M.}~\bibnamefont
  {Ritsch-Marte}},\ }\bibfield  {title} {\bibinfo {title} {Structured light for
  microscopy in \textit{ {R}oadmap on structured light}, {H}alina
  {R}ubinsztein-{D}unlop et al.},\ }\bibfield  {journal} {\bibinfo  {journal}
  {J. Opt}\ }\textbf {\bibinfo {volume} {19}},\ \href
  {https://doi.org/10.1088/2040-8978/19/1/013001}
  {10.1088/2040-8978/19/1/013001} (\bibinfo {year} {2017})\BibitemShut
  {NoStop}%
\bibitem [{\citenamefont {M\"uller}\ and\ \citenamefont
  {Enderlein}(2010)}]{ImageScanningMicroscopy}%
  \BibitemOpen
  \bibfield  {author} {\bibinfo {author} {\bibfnamefont {C.~B.}\ \bibnamefont
  {M\"uller}}\ and\ \bibinfo {author} {\bibfnamefont {J.}~\bibnamefont
  {Enderlein}},\ }\bibfield  {title} {\bibinfo {title} {Image scanning
  microscopy},\ }\href {https://doi.org/10.1103/PhysRevLett.104.198101}
  {\bibfield  {journal} {\bibinfo  {journal} {Phys. Rev. Lett.}\ }\textbf
  {\bibinfo {volume} {104}},\ \bibinfo {pages} {198101} (\bibinfo {year}
  {2010})}\BibitemShut {NoStop}%
\bibitem [{\citenamefont {Aleksanyan}\ \emph {et~al.}(2017)\citenamefont
  {Aleksanyan}, \citenamefont {Kravets},\ and\ \citenamefont
  {Brasselet}}]{CoronaLightValve}%
  \BibitemOpen
  \bibfield  {author} {\bibinfo {author} {\bibfnamefont {A.}~\bibnamefont
  {Aleksanyan}}, \bibinfo {author} {\bibfnamefont {N.}~\bibnamefont
  {Kravets}},\ and\ \bibinfo {author} {\bibfnamefont {E.}~\bibnamefont
  {Brasselet}},\ }\bibfield  {title} {\bibinfo {title} {Multiple-star system
  adaptive vortex coronagraphy using a liquid crystal light valve},\ }\href
  {https://doi.org/10.1103/PhysRevLett.118.203902} {\bibfield  {journal}
  {\bibinfo  {journal} {Phys. Rev. Lett.}\ }\textbf {\bibinfo {volume} {118}},\
  \bibinfo {pages} {203902} (\bibinfo {year} {2017})}\BibitemShut {NoStop}%
\bibitem [{\citenamefont {Dennis}\ \emph {et~al.}(2009)\citenamefont {Dennis},
  \citenamefont {O'Holleran},\ and\ \citenamefont {Padgett}}]{VorticesDennis}%
  \BibitemOpen
  \bibfield  {author} {\bibinfo {author} {\bibfnamefont {M.~R.}\ \bibnamefont
  {Dennis}}, \bibinfo {author} {\bibfnamefont {K.}~\bibnamefont {O'Holleran}},\
  and\ \bibinfo {author} {\bibfnamefont {M.~J.}\ \bibnamefont {Padgett}},\
  }\bibfield  {title} {\bibinfo {title} {Chapter 5 singular optics: Optical
  vortices and polarization singularities}\ }(\bibinfo  {publisher}
  {Elsevier},\ \bibinfo {year} {2009})\ pp.\ \bibinfo {pages}
  {293--363}\BibitemShut {NoStop}%
\bibitem [{\citenamefont {Courtial}\ \emph {et~al.}(1998)\citenamefont
  {Courtial}, \citenamefont {Robertson}, \citenamefont {Dholakia},
  \citenamefont {Allen},\ and\ \citenamefont {Padgett}}]{RotFreqShiftPadgett}%
  \BibitemOpen
  \bibfield  {author} {\bibinfo {author} {\bibfnamefont {J.}~\bibnamefont
  {Courtial}}, \bibinfo {author} {\bibfnamefont {D.~A.}\ \bibnamefont
  {Robertson}}, \bibinfo {author} {\bibfnamefont {K.}~\bibnamefont {Dholakia}},
  \bibinfo {author} {\bibfnamefont {L.}~\bibnamefont {Allen}},\ and\ \bibinfo
  {author} {\bibfnamefont {M.~J.}\ \bibnamefont {Padgett}},\ }\bibfield
  {title} {\bibinfo {title} {Rotational frequency shift of a light beam},\
  }\href {https://doi.org/10.1103/PhysRevLett.81.4828} {\bibfield  {journal}
  {\bibinfo  {journal} {Phys. Rev. Lett.}\ }\textbf {\bibinfo {volume} {81}},\
  \bibinfo {pages} {4828} (\bibinfo {year} {1998})}\BibitemShut {NoStop}%
\bibitem [{\citenamefont {Schechner}\ \emph {et~al.}(1996)\citenamefont
  {Schechner}, \citenamefont {Piestun},\ and\ \citenamefont
  {Shamir}}]{RotPropPiestun}%
  \BibitemOpen
  \bibfield  {author} {\bibinfo {author} {\bibfnamefont {Y.~Y.}\ \bibnamefont
  {Schechner}}, \bibinfo {author} {\bibfnamefont {R.}~\bibnamefont {Piestun}},\
  and\ \bibinfo {author} {\bibfnamefont {J.}~\bibnamefont {Shamir}},\
  }\bibfield  {title} {\bibinfo {title} {Wave propagation with rotating
  intensity distributions},\ }\href {https://doi.org/10.1103/PhysRevE.54.R50}
  {\bibfield  {journal} {\bibinfo  {journal} {Phys. Rev. E}\ }\textbf {\bibinfo
  {volume} {54}},\ \bibinfo {pages} {R50} (\bibinfo {year} {1996})}\BibitemShut
  {NoStop}%
\bibitem [{\citenamefont {Turunen}\ and\ \citenamefont
  {Friberg}(2010)}]{InvariantBeams}%
  \BibitemOpen
  \bibfield  {author} {\bibinfo {author} {\bibfnamefont {J.}~\bibnamefont
  {Turunen}}\ and\ \bibinfo {author} {\bibfnamefont {A.~T.}\ \bibnamefont
  {Friberg}},\ }\bibfield  {title} {\bibinfo {title} {Propagation-invariant
  optical fields},\ }\href
  {https://doi.org/https://doi.org/10.1016/S0079-6638(10)05406-5.} {\bibfield
  {journal} {\bibinfo  {journal} {Progress in Optics}\ }\textbf {\bibinfo
  {volume} {54}},\ \bibinfo {pages} {1} (\bibinfo {year} {2010})}\BibitemShut
  {NoStop}%
\bibitem [{\citenamefont {Coullet}\ \emph {et~al.}(1989)\citenamefont
  {Coullet}, \citenamefont {Gil},\ and\ \citenamefont {F.}}]{OpticalVortices}%
  \BibitemOpen
  \bibfield  {author} {\bibinfo {author} {\bibfnamefont {P.}~\bibnamefont
  {Coullet}}, \bibinfo {author} {\bibfnamefont {F.}~\bibnamefont {Gil}},\ and\
  \bibinfo {author} {\bibfnamefont {R.}~\bibnamefont {F.}},\ }\bibfield
  {title} {\bibinfo {title} {Optical vortices},\ }\href@noop {} {\bibfield
  {journal} {\bibinfo  {journal} {Optics Communications}\ }\textbf {\bibinfo
  {volume} {73}},\ \bibinfo {pages} {403} (\bibinfo {year} {1989})}\BibitemShut
  {NoStop}%
\bibitem [{\citenamefont {W.Goodman}(2005)}]{FourierO}%
  \BibitemOpen
  \bibfield  {author} {\bibinfo {author} {\bibfnamefont {J.}~\bibnamefont
  {W.Goodman}},\ }\href@noop {} {\emph {\bibinfo {title} {Introduction to
  Fourier Optics}}},\ \bibinfo {edition} {3rd}\ ed.\ (\bibinfo  {publisher}
  {Roberts \& Company},\ \bibinfo {year} {2005})\BibitemShut {NoStop}%
\bibitem [{\citenamefont {Lehtinen}(2020)}]{FormulaireBessel}%
  \BibitemOpen
  \bibfield  {author} {\bibinfo {author} {\bibfnamefont {N.~G.}\ \bibnamefont
  {Lehtinen}},\ }\href
  {http://nlpc.stanford.edu/nleht/Science/reference/bessel.pdf} {\bibinfo
  {title} {Everything a physicist needs to know about bessel functions}}
  (\bibinfo {year} {2020})\BibitemShut {NoStop}%
\bibitem [{\citenamefont {Zheng}\ \emph {et~al.}(2013)\citenamefont {Zheng},
  \citenamefont {Horstmeyer},\ and\ \citenamefont {Yang}}]{PtychoFourier}%
  \BibitemOpen
  \bibfield  {author} {\bibinfo {author} {\bibfnamefont {G.}~\bibnamefont
  {Zheng}}, \bibinfo {author} {\bibfnamefont {R.}~\bibnamefont {Horstmeyer}},\
  and\ \bibinfo {author} {\bibfnamefont {C.}~\bibnamefont {Yang}},\ }\bibfield
  {title} {\bibinfo {title} {Wide-field, high-resolution fourier ptychographic
  microscopy},\ }\href@noop {} {\bibfield  {journal} {\bibinfo  {journal}
  {Nature Photon.}\ }\textbf {\bibinfo {volume} {7}},\ \bibinfo {pages} {739}
  (\bibinfo {year} {2013})}\BibitemShut {NoStop}%
\bibitem [{\citenamefont {Ruh}\ \emph {et~al.}(2018)\citenamefont {Ruh},
  \citenamefont {Mutschler}, \citenamefont {Michelbach},\ and\ \citenamefont
  {Rohrbach}}]{ROCSmicroRohrbach}%
  \BibitemOpen
  \bibfield  {author} {\bibinfo {author} {\bibfnamefont {D.}~\bibnamefont
  {Ruh}}, \bibinfo {author} {\bibfnamefont {J.}~\bibnamefont {Mutschler}},
  \bibinfo {author} {\bibfnamefont {M.}~\bibnamefont {Michelbach}},\ and\
  \bibinfo {author} {\bibfnamefont {A.}~\bibnamefont {Rohrbach}},\ }\bibfield
  {title} {\bibinfo {title} {Superior contrast and resolution by image
  formation in rotating coherent scattering ({ROCS}) microscopy},\ }\href
  {https://doi.org/10.1364/OPTICA.5.001371} {\bibfield  {journal} {\bibinfo
  {journal} {Optica}\ }\textbf {\bibinfo {volume} {5}},\ \bibinfo {pages}
  {1371} (\bibinfo {year} {2018})}\BibitemShut {NoStop}%
\bibitem [{\citenamefont {O.Haeberlé}\ \emph {et~al.}(2010)\citenamefont
  {O.Haeberlé}, \citenamefont {K.Belkebir}, \citenamefont {H.Giovaninni},\
  and\ \citenamefont {A.Sentenac}}]{Tomoreview}%
  \BibitemOpen
  \bibfield  {author} {\bibinfo {author} {\bibnamefont {O.Haeberlé}}, \bibinfo
  {author} {\bibnamefont {K.Belkebir}}, \bibinfo {author} {\bibnamefont
  {H.Giovaninni}},\ and\ \bibinfo {author} {\bibnamefont {A.Sentenac}},\
  }\bibfield  {title} {\bibinfo {title} {Tomographic diffractive microscopy:
  basics, techniques and perspectives},\ }\href@noop {} {\bibfield  {journal}
  {\bibinfo  {journal} {Journal of Modern Optics}\ }\textbf {\bibinfo {volume}
  {57}},\ \bibinfo {pages} {686 } (\bibinfo {year} {2010})}\BibitemShut
  {NoStop}%
\bibitem [{\citenamefont {Young}\ and\ \citenamefont {Kukura}(2019)}]{iSCAT}%
  \BibitemOpen
  \bibfield  {author} {\bibinfo {author} {\bibfnamefont {G.}~\bibnamefont
  {Young}}\ and\ \bibinfo {author} {\bibfnamefont {P.}~\bibnamefont {Kukura}},\
  }\bibfield  {title} {\bibinfo {title} {Interferometric scattering
  microscopy},\ }\href {https://doi.org/10.1146/annurev-physchem-050317-021247}
  {\bibfield  {journal} {\bibinfo  {journal} {Annual Review of Physical
  Chemistry}\ }\textbf {\bibinfo {volume} {70}},\ \bibinfo {pages} {301}
  (\bibinfo {year} {2019})}\BibitemShut {NoStop}%
\bibitem [{\citenamefont {Popoff}(2016)}]{DMDsetting}%
  \BibitemOpen
  \bibfield  {author} {\bibinfo {author} {\bibfnamefont {S.}~\bibnamefont
  {Popoff}},\ }\bibfield  {title} {\bibinfo {title} {Setting up a {DMD}:
  Diffraction effects},\ }\href
  {http://wavefrontshaping.net/index.php/component/content/article/57-community/tutorials/spatial-lights-modulators-slms/131-setting-up-a-dmd-diffraction-effects}
  {\bibfield  {journal} {\bibinfo  {journal} {www.wavefrontshaping.net}\ }
  (\bibinfo {year} {2016})}\BibitemShut {NoStop}%
\bibitem [{\citenamefont {Popoff}(2019)}]{DMDab}%
  \BibitemOpen
  \bibfield  {author} {\bibinfo {author} {\bibfnamefont {S.}~\bibnamefont
  {Popoff}},\ }\bibfield  {title} {\bibinfo {title} {Setting up a dmd/slm:
  Aberration effects},\ }\href {https://www.wavefrontshaping.net/post/id/23}
  {\bibfield  {journal} {\bibinfo  {journal} {www.wavefrontshaping.net}\ }
  (\bibinfo {year} {2019})}\BibitemShut {NoStop}%
\end{thebibliography}%

\onecolumngrid

\clearpage
\section*{Supplementary informations}

\tableofcontents
\clearpage

\section{Concept}

\begin{figure}[h!] 
    \begin{center}
        \includegraphics[width=0.8\columnwidth]{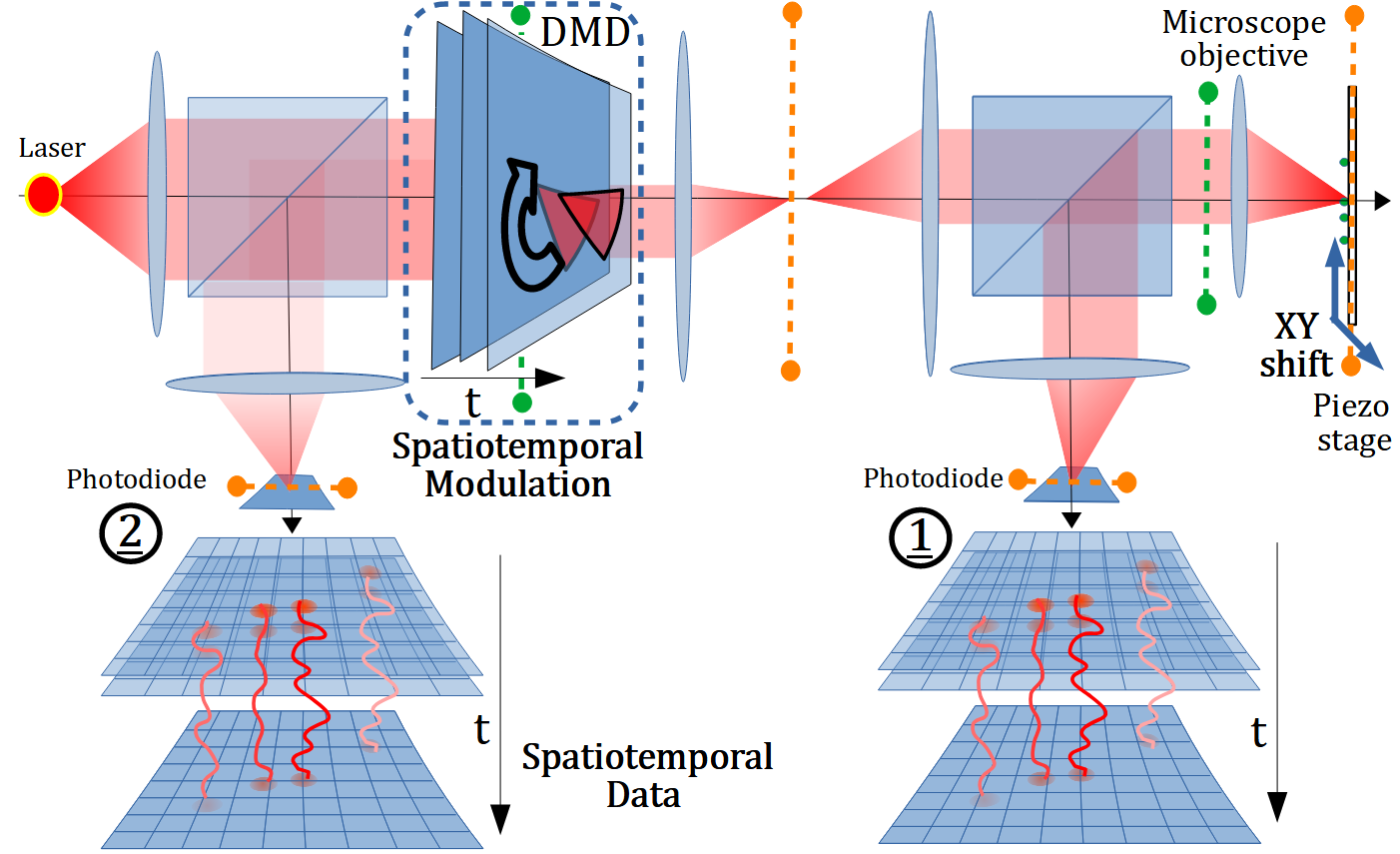}
        \caption{\label{fig:ConfocalDMD2phot} \small{\textbf{Schematic view of the experimental setup.} The DMD, depicted in transmission here for the sake of simplicity, is conjugated to the back focal plane of the microscope objective. Conjugated planes are shown using dotted lines of the same colors (orange or green). To pass from a plane indicated by one color to a plane indicated by the other color, a 2D Fourier transform plus a scaling operation are performed. Eventually, a filtering operation by the pupil's function is also achieved as it is the case for the high spatial frequency content filtered by the microscope objective.}}
    \end{center}
\end{figure}

Contrarily to the main text, we consider a more complete version of the setup shown on Figure~\ref{fig:ConfocalDMD2phot}. With this setup, the backscattered field is collected using two different photodiodes with or without a second pass on the DMD (referred respectively as photodiode 1 and 2). This is achieved by adding or suppressing a polarizing beamsplitter cube behind the microscope objective. This corresponds more accurately to the real experimental setup and allows for a more profound physical understanding. In the main text, only the simulation and experimental results corresponding to the second photodiode are presented since they are the most \textit{interesting results}.

\section{Physical \& Mathematical model}

The goal of this section is to compute the point-spread function (PSF) and the coherent transfer function (the monochromatic optical transfer function, thereafter referred as the CTF) of the system.

All the calculations are made for the field and not the intensity except in the part~\ref{subs:Comp} \textit{'Observation of vortex-like images with intensity-only measurements'} concerning the optical sectioning where the coherent transfer function is explicitly calculated in intensity. This is reminded with the 'I' of ICTF (for \textit{Intensity Coherent Transfer Function}). 

\subsection{Model for the confocal microscope}

The model used here is standard, it is presented in details in~\cite{JMertz}. Under the hypothesis of a punctual detector the PSF is the product of the illumination and collection PSFs. In the Fourier domain it becomes a convolution :

\begin{equation}\label{PSFconf} 
    \textrm{PSF}_{\textrm{conf}}=\textrm{PSF}_{\textrm{ill}}\cdot \textrm{PSF}_{\textrm{coll}} \xLeftrightarrow{\large{TF_{2D}}} \textrm{CTF}_{\textrm{conf}}=\textrm{CTF}_{\textrm{ill}}\otimes \textrm{CTF}_{\textrm{coll}}
\end{equation}

\subsection{Influence of the pinhole size on the PSF}

In practice, the pinhole size cannot be too small in order to collect a significative photon flux. To understand the effect of this parameter it is possible to examine two limiting cases: in addition to the case of a punctual pinhole (which is the confocal case) we consider the case of an infinite pinhole. In this second case, the sample is illuminated on a single diffraction-limited point and all the light from the sample is collected in a full-field way. What limits the resolution is then the illumination. Thus, in this limiting case, the PSF is the illumination PSF equivalent to the full-field PSF namely PSF$_{\!f\!f}$. The convolution comes from the XY scanning of the sample. Thus, in a symmetrical configuration, the PSF evolves from PSF$_{\!f\!f}$ to PSF$_{\!f\!f}^2$. Hence, the lateral resolution is always slightly higher than in the full-field configuration.

\subsection{Field created by a rotating surface element on the DMD}

\begin{figure}[t!] 
    \begin{center}
        \includegraphics[width=0.8\columnwidth]{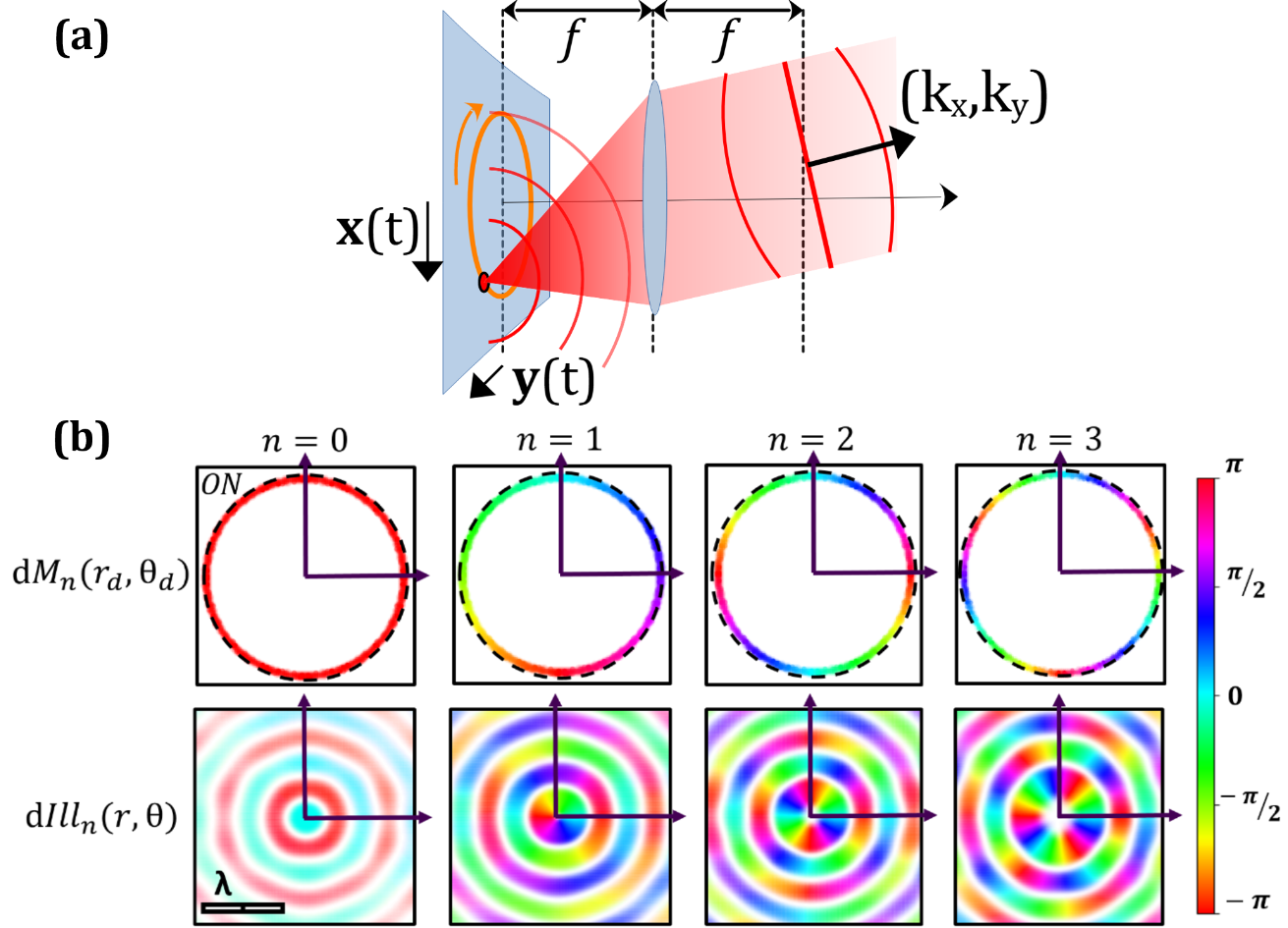}
        \caption{\label{fig:ExpPF_dM_Ill} \small{\textbf{Situation of a rotating point-source in the plane of DMD and associated fields} obtained by numerical simulation. \textbf{a)} Schematic view of the experimental configuration : a rotating point is displayed on the DMD, it corresponds to a plane wave in the focal plane. \textbf{b)} Representation of the first term of the Fourier series $dM_n$ and $Ill_n$  for $n \in [\![0;3]\!]$ in weighted phase representation for $r_d=R$. }}
    \end{center}
\end{figure}

First, let us consider the case where the pattern $dM$ on the DMD is a single rotating point at position $\overrightarrow{r_d}(t)=\left(x_d(t),y_d(t)\right)=\left(r_d \cos(\Omega t) , r_d \sin(\Omega t)\right)$ equivalent to $(r_d,\theta_d(t))=(r_d,\Omega t)$ receiving a plane wave directed by the optical axis. This point behaves a point-source as shown in Figure \ref{fig:ExpPF_dM_Ill}.\textbf{a)}. In cylindrical coordinates it writes $\textrm{d}M(\overrightarrow{r},t | \overrightarrow{r_d})=\delta(r-r_d)\cdot \delta(\theta-\theta_d) e^{i\omega_0 t}$. As it is a time-periodic pattern it can be decomposed as a Fourier series:
\begin{equation} \label{dMn}
    \textrm{d}M(\overrightarrow{r},t | \overrightarrow{r_d})=\sum \limits_{n=-\infty}^{\infty} \delta(r-r_d) e^{in(\theta-\theta_d)}e^{in\Omega t} \cdot e^{i\omega_0 t}
\end{equation}

In the plane of the modulator the field already writes as a frequency comb and the field is already a sum of vortex.  The phase of the vortices is in fact an encoding of the angular position of the moving point in time. This first theoretical result is also retrieved by numerical simulation as shown in Figure~\ref{fig:ExpPF_dM_Ill}.\textbf{b}.

This kind of beams with vortex singularities are well-known in optics~\cite{VorticesDennis}. This effect, which can be understood as a consequence of the Doppler effect, has already been highlighted in the litterature~\cite{RotFreqShiftPadgett, RotPropPiestun} but has not yet applied to microscopy to our knowledge. 

Due to their topological nature, the vortex structure is conserved by propagation and by spatial Fourier transform~\cite{InvariantBeams,OpticalVortices}.
By applying the principles of Fourier optics~\cite{FourierO}, the field in the focal plane of the microscope objective associated to a moving point is a plane wave of wavevector $\overrightarrow{k}(t)=\left(\frac{2\pi r_d \cos(\Omega t)}{\lambda f}, \frac{2\pi r_d \sin(\Omega t)}{\lambda f}\right)$ as shown in Figure \ref{fig:ExpPF_dM_Ill}.
The \textit{elementary pattern} displayed $\dd M(r,\theta,t | r_d,\theta_d)=\delta(r-r_d)\delta(\theta-(\theta_d+\Omega t))$ is a point-source at position $(r_d,\theta_d)$ in the plane of the DMD. We wish to express the field in the sample's plane at position $(r,\theta)$ which is in the Fourier plane of the DMD. The basic principles of Fourier optics~\cite{FourierO} states that the image of a point is a plane wave (within the \textit{exit pupil}, this is neglected here since the field is considered near the optical axis). With cylindrical coordinates and in the sample's plane $\left(\overrightarrow{u_r},\overrightarrow{u_\theta}\right)$, the wavevector associated to $\dd M$ writes $\overrightarrow{k}(r_d,\theta_d)= \frac{2 \pi r_d}{\lambda f} \overrightarrow{u_r} + (\theta_d+\Omega t )\overrightarrow{u_\theta} $ where the carrier frequency $\omega_0$ is neglected. Using trigonometric properties, the elementary illumination $\dd Ill$ associated to $\dd M$ can write:
\begin{equation} \label{eq:dIll}
    \dd Ill(r,\theta,t|r_d,\theta_d)=e^{\frac{2\pi i r r_d}{\lambda f}(\cos(\theta)\cos(\theta_d+\Omega t)+\sin(\theta)\sin(\theta_d+\Omega t))} \propto e^{\frac{2\pi i r_d r }{\lambda f}\sin(\Omega t+\theta_d+\theta)} \ \ .
\end{equation}

\noindent Now, from~\cite{FormulaireBessel}, we use the generating function of Bessel's functions: 
\begin{equation}\label{eq:GenBessel}
g(x,t)=e^{\frac{x}{2}(z-\frac{1}{z})}=\sum_{n=-\infty}^{\infty}J_n(x)z^n
\end{equation}
by identifying $z=e^{i(\theta_d+\theta+\Omega t)}$ and $x=\frac{2 \pi r_d r}{\lambda f}$.
By linearity, each term of the sum thereafter named $\dd Ill_n$ is the 2D Fourier transform of each term from Equation~(\ref{dMn}) with a scaling factor (same calculations as the alternative demonstration below). This expression is equivalent to the field emitted by a rotating receiver in the acoustic experiment of~\cite{PRAppliedAcoustic_Noetinger} with an additional phase shift between each term. As in the acoustic experiment, the field is again a frequency comb of spacing $\Omega$ centered on $\omega_0$. Also, for each frequency, the field is vortex-like and concentrated on rings whose radius increase with $n$.
This is easily confirmed numerically by performing the 2D Fourier transform of a sequence with a rotating point and then computing the first terms of the temporal Fourier series as illustrated in Figure~\ref{fig:ExpPF_dM_Ill}. We retrieve the results from the acoustical experiment.

Another option is to decompose from the beginning the field on the harmonics $n\Omega$ as a Fourier series $\dd~M(t)~=~\sum \limits_{n=-\infty}^{\infty} \dd M_n e^{in\Omega t}$ with $n \in \mathbb{Z}$, where we need to compute the Fourier coefficients defined by:
\begin{equation} \label{eq:d_cn}
     \dd M_n(r_{D},\theta_{D})=\frac{1}{T}\int_0^{T}M(r_{D},\theta_{D},t)e^{-i n\Omega t} \dd  t
\end{equation}
Then a possible definition of $J_n(x)$ shows up :
\begin{equation} \label{eq:DefBesseln}
J_n(x)=\int_0^{2\pi}e^{i x \sin(\theta)-i n \theta }\dd \theta
\end{equation}

\subsection{Field created by a rotating sector and alternative demonstration}

To provide another proof of this formula let us consider the example of a sector rotating on the DMD of angular width $\alpha$ and radial width $R$ as shown on figure~\ref{fig:MotifCalc}. It can be written as a product of two gate functions in radius-angle $(r,\theta)$ or in radius-time $(r,t)$ since the two variables are now linked:
\begin{figure}[t!] 
    \begin{center}
        \includegraphics[width=0.9\columnwidth]{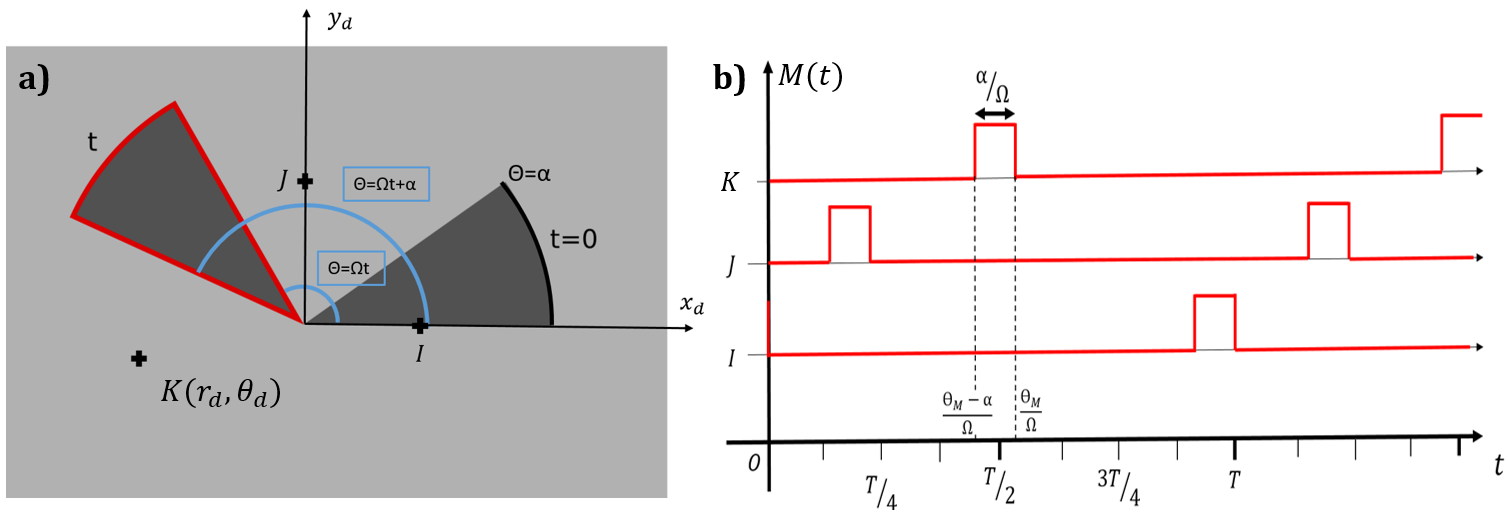}
        \caption{\label{fig:MotifCalc} \small{\textbf{Rotating sector displayed on the DMD} \textbf{a)} Scheme of the DMD seen at time $t$, comparison with $t=0$. \textbf{b)} Chronogram of the signal at different points on the DMD surface.}}
    \end{center}
\end{figure}

\begin{equation}
    M(r_{D},\theta_{D},t) \propto \Pi_{[\Omega t,\Omega t+\alpha]}(\theta_d)\cdot\Pi_{[0,R]}(r)= \Pi_{\left[\frac{\theta_d}{\Omega},\frac{\theta_d+\alpha}{\Omega}\right]}(t)\cdot\Pi_{[0,R]}(r)
\end{equation}
The Fourier coefficients write: 
\begin{align*}
    M_n(r_{D},\theta_{D})
    &=\frac{1}{T}\int_0^{2\pi}\Pi_{\left[\frac{\theta_d}{\Omega},\frac{\theta_d+\alpha}{\Omega}\right]}(t)\Pi_{[0,R]}(r_d)e^{-i\Omega t}(r)\dd t
    \\
    &=\frac{1}{T}\Pi_{[0,R]}(r_d)\int_{\frac{\theta_d}{\Omega}}^{\frac{\theta_d+\alpha}{\Omega}}e^{-in\Omega t}\dd t
    \\
    &=\frac{ \alpha}{2\pi}\textrm{sinc}\left(\frac{n\alpha}{2}\right)e^{-in(\theta_d +\alpha/2)}\Pi_{[0,R]}
\end{align*}
To obtain the field in the object plane the 2D Fourier transform writes:
\begin{align}
        Ill(r,\theta,t)&=\iint \limits_{-\infty}^{\ \ \ +\infty}M(r_d,\theta_d,t)e^{-i\frac{2\pi}{\lambda f}(r_d r\cdot (cos(\theta_d)cos(\theta)+sin(\theta_d)sin(\theta))}r_d \dd r_d \dd \theta_d \\
        &=\iint \limits_{-\infty}^{\ \ \ +\infty}M(r_d,\theta_d,t)e^{-i\frac{2\pi}{\lambda f}r_d r\cdot cos(\theta_d-\theta)}r_d \dd r_d \dd\theta_d \\
        &\propto \iint \limits_{0,0}^{\ \ \ R,2\pi}M(r_d,\theta_d,t)e^{-i\frac{2\pi}{\lambda f}r_d r\cdot \sin(\theta_d-\theta+\pi)}r_d \dd r_d \dd \theta_d \\
        &=\int \limits_{0}^{+\infty}\Pi_{[0,R]}(r)\int \limits_{0}^{2\pi}\sum \limits_{n=-\infty}^{\infty} \textrm{sinc}\left(\frac{n\alpha}{2}\right)e^{-in(\theta_d +\frac{\alpha}{2})} e^{-i\frac{2\pi}{\lambda f}r_d r\cdot \sin(\theta_d-\theta+\pi)}d\theta_d r_d dr_d \ e^{in\Omega t}
\end{align}

\noindent Using the definition of Bessel function from above~(Equation~(\ref{eq:DefBesseln})):
\begin{equation}
     Ill(r,\theta,t)\propto i^n \textrm{sinc}\left(\frac{n\alpha}{2}\right)e^{in(\frac{\alpha}{2}+\pi/2+\theta)}\int \limits_{0}^{R} J_n\left(\frac{2 \pi}{\lambda f}r_D r\right) r_d dr_d
\end{equation}
Applying the formula \ref{PSFconf} gives the two PSFs:
\begin{align}
    \textrm{PSF}_{1}(r,\theta,t) &\propto Ill(r,\theta,t)\cdot \frac{J_{1}\left(\frac{2\pi ON }{\lambda}r\right)}{r} \label{PSF1eq}\\
   \textrm{PSF}_{2}(r,\theta,t) &\propto Ill(r,\theta,t)\cdot Ill(r,\theta+\pi,t) \label{PSF2eq} \ \ .
\end{align}
To generalize to any pattern, it is possible to deform continuously this specific pattern to obtain a surface element of angular and radial width $\dd \theta_d$ and $\dd r_d$ then we transform :
\begin{equation}
    \small{\begin{pmatrix} R  \\r_{min} \\ \alpha/2+\pi/2 \\\alpha \end{pmatrix} \to \begin{pmatrix} r_d+dr_d  \\ r_d \\ \theta_d \\ d\theta_d \end{pmatrix}}
\end{equation}
The two integrals \textit{disappear} leaving only:
\begin{equation}
    \dd Ill_n(r,\theta) = i^n e^{-in(\theta_d+\theta)} J_n\left(\frac{2 \pi r_D}{\lambda f}r\right)r_d \dd r_d \dd\theta_d 
\end{equation}
It is then possible to integrate this expression on the DMD surface to find the PSFs.
\subsection{Field created by any rotating pattern}

In order to acquire an image, it is recommended to concentrate energy in the object's plane near the optical axis. This can be achieved by summing waves constructively, a process better known as \textit{focusing}. This is achieved here by displaying a larger pattern on the DMD. Moreover, this allows to take advantage of the high number of degrees of freedom available with the DMD. By integrating the contribution of each surface elements $\dd M$ which is equivalent to summing the angular spectrum, the field in the focal plane \textit{ie.} the illumination $Ill$ associated to the pattern $M(r_d,\theta_d)$ is computed:
\begin{equation} \label{s_pf_tot}
    Ill(r,\theta,t)=\iint \limits_{M} e^{i \overrightarrow{k}\cdot \overrightarrow{r}}= \sum \limits_{n=-\infty}^{\infty} i^n \iint \limits_{M(r_d,\theta_d)} J_n \left(\frac{2 \pi}{\lambda f}r_d r \right) \cdot e^{in\theta}  e^{in\Omega t} \cdot e^{i\omega_0 t}  r_d \dd r_d \dd \theta_d
\end{equation}
To express the confocal images built for each photodiode it is necessary to take into account the collection of the backscattered wave by the objective and possible filtering during the second passage by the DMD. For the sake of simplicity we express the PSF for each photodiode in terms of filed and not in intensity so that the images are given by:
\begin{equation} \label{Im}
    Im_i= |\textrm{PSF}_i \otimes Obj|^2
\end{equation}

where $Obj$ is the object in the focal plane.
For the first photodiode, the backscattered light is collected with all the numerical aperture so $\textrm{PSF}_{\textrm{coll}}$ is just the full-field PSF \textit{ie.} the Airy function:
\begin{equation}\label{PSF1}
    \textrm{PSF}_{1}(r,\theta,t) \propto Ill(r,\theta,t)\cdot \frac{J_{1}\left(\frac{2\pi ON }{\lambda}r\right)}{r}
\end{equation}
This configuration seems close to Fourier ptychography~\cite{PtychoFourier} or ROCS microscopy ~\cite{ROCSmicroRohrbach}, the difference lies in the confocal configuration : we use a bucket point-like receptor and a scanning process.
For the second photodiode only the wavevectors associated to the pattern are collected. The reflection on the sample has to be taken into account, in this case all the incident wavevectors are mirrored after reflection so that:
\begin{equation}\label{PSF2}
    \textrm{PSF}_{2}(r,\theta,t) \propto Ill(r,\theta,t)\cdot Ill(r,\theta+\pi,t) \ \ .
\end{equation}
As a periodic function of time, these PSFs are decomposed using Fourier series as a sum of PSFs for each frequency $\omega_0+n \Omega$ named the \textit{harmonic PSFs} thereafter $n$-indexed and written $\textrm{PSF}_{i,n}$ (with $i$ being 1 or 2).

\subsection{Example : PSF associated to a 45$^\circ$ rotating sector}

\begin{figure}[t!]
    \begin{center}
        \includegraphics[width=0.7\columnwidth]{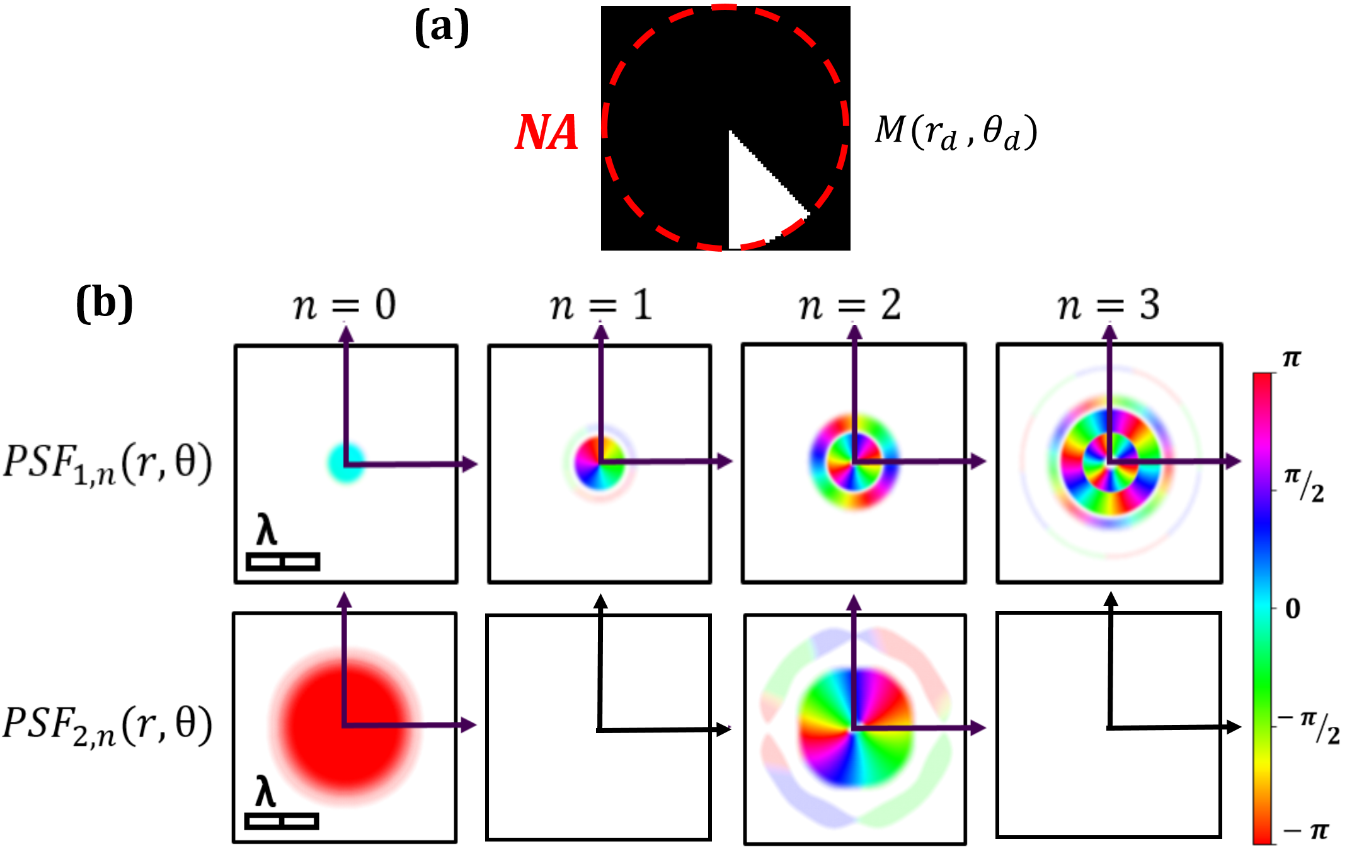}
        \caption{\label{fig:Rot45degPSF} \small{\textbf{PSF associated to a given pattern.}} \textbf{a)} The rotating pattern is a truncated 45° sector in order to collect only high spatial frequencies. It is shown here next to the numerical aperture ($N\!A$) of the microscope objective. \textbf{b)} Representation of the PSF on each photodiode with and without a second pass on the DMD respectively $PSF_{1,n}$ and $\textrm{PSF}_{2,n}$ for $n \in [\![0;3]\!]$. }
    \end{center}
\end{figure}

To illustrate the previous expresssions, we choose to consider, as a rotating pattern for the DMD, a 45° sector (Figure \ref{fig:Rot45degPSF}.\textbf{a)}). The $\textrm{PSF}_{1,n}$ are also vortices of similar size to the confocal PSF localized next to the focal point and with less secondary lobes than the illumination $Ill_n$ thanks to the focusing of the backscattered wave by the microscope objective. 
The PSF for the second photodiode involves a product of two frequency combs. This is another frequency comb whose coefficients can be computed using Cauchy product but is rather difficult to interpret. Therefore it is more straightforward to use numerical simulation results. As a consequence of the difference between the acoustical and optical experiments mentioned in the main text consisting of a spherical phase disappearing in the optical experiment $\textrm{PSF}_{2,n}$ are null if $n$ is odd and are significantly larger than the confocal PSF otherwise as seen on Figure~\ref{fig:Rot45degPSF}.\textbf{a)}. With this kind of PSF it is not possible to enhance the resolution of the confocal microscope

\subsection{Explanations on the resolution decrease after a second pass on the DMD} \label{subs:Explations}

\begin{figure}[t!] 
    \begin{center}
        \includegraphics[width=0.7\columnwidth]{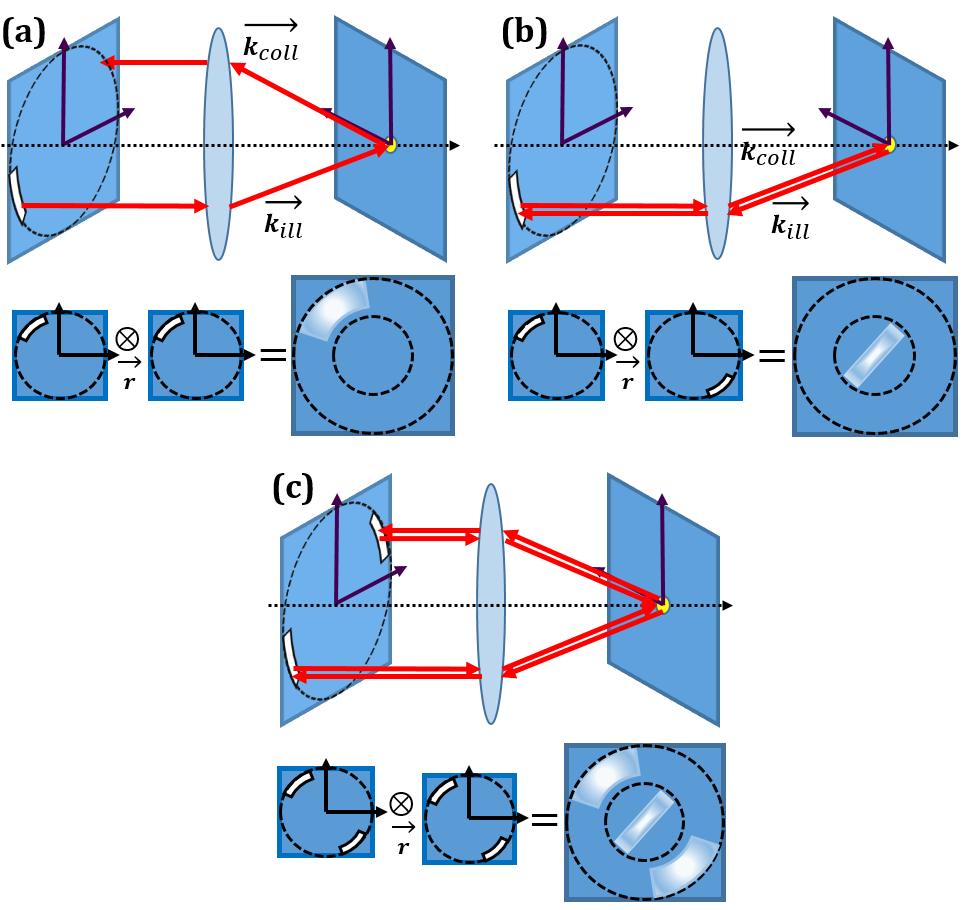}
        \caption{\label{fig:ConfigOTF} \small{\textbf{Different patterns of the DMD with the associated wavevector in the sample planes and CTF } \textbf{a)} Hypothetic configuration neglecting the reflection on the sample allowing to collect high frequencies component of the sample. \textbf{b)} Real configuration and associated CTF. \textbf{c)} Symmetrization of the pattern. High and low frequencies are collected.}}
    \end{center}
\end{figure}

To understand the low resolution obtained with the 45$^\circ$ sector (see Figure 3 of the main text) let us consider the configurations depicted by Figure \ref{fig:ConfigOTF}.\\
Using a limited zone of the DMD that we consider to be point-like at the edge of the pupil, a plane wave is projected onto the sample. If the sample consists in a simple point-like object then it backscatters a wave in phase with the plane wave. The collected wave corresponds to the same plane wave. Thus, the phase law applied in the first place is exactly compensated during the collection: for each position of the mechanical scan the same signal is collected. This does not allow to build an image since the signal collected is always the same. In order to collect spatial frequencies close to $\frac{2N\!A}{\lambda}$ the angular spectrum directed by the symmetrical plane wave should be collected. This can be achieved by doubling the pattern. Then the high frequencies are collected but the low frequencies are also collected.

\begin{figure}[b!] 
    \begin{center}
        \includegraphics[width=0.9\columnwidth]{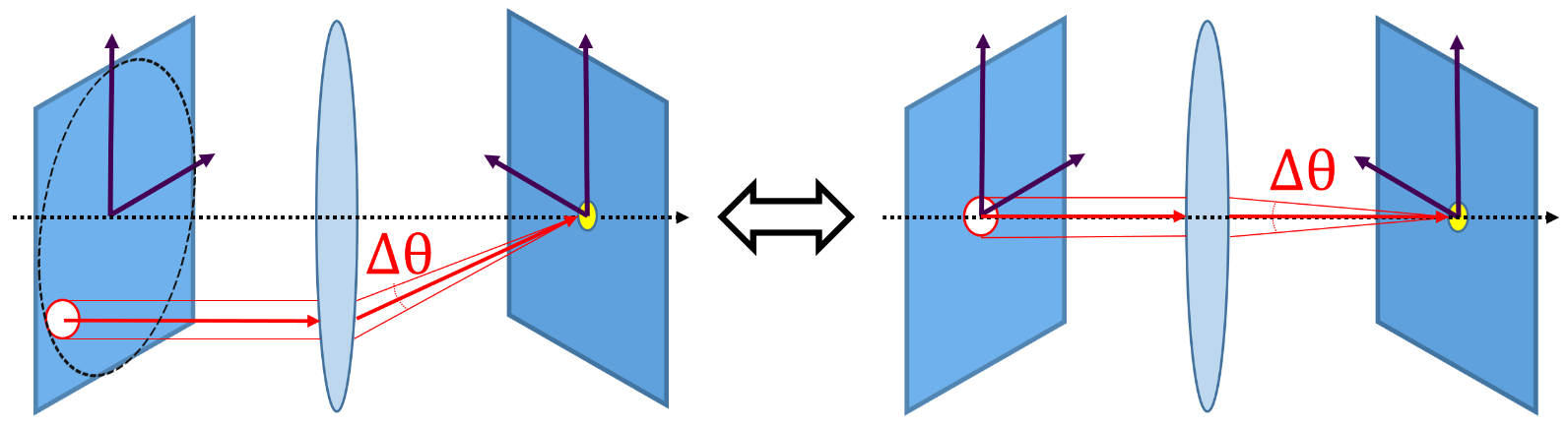}
        \caption{\label{fig:ConfigON} \small{\textbf{Equivalence between the illumination associated with off-axis and centered patterns.}  The illuminations differs only by a phase ramp. Then the image that can be built by a DMD pattern defining a cone of angular width $\Delta \theta$ is of resolution defined by the numerical aperture $n \sin(\Delta \theta)$ as if the pattern was centered on the optical axis. } }
    \end{center}
\end{figure}

The results obtained in acoustics~\cite{PRAppliedAcoustic_Noetinger} and the previous calculations are obtained in field only. Accessing the field in optics requires an interferometric setup such as phase-shifting or off-axis holography and adds experimental constraints. As we use an intensity detector here, the quality of the image depends on the interference of all the wave collected. Using a reference beam as in tomographic~\cite{Tomoreview} or iSCAT microscopy~\cite{iSCAT} we could retrieve spatial features limited by $2N\!A/\lambda$ but we chose here to collect only light from the sample as a reference beam would not be stable enough for long acquisition time as in confocal scanning. In this case, a collection wavevector with an important angle must be collected from the sample. In the full-field configuration they are in the cone of angular width $\theta$ (with $n\sin(\theta)=N\!A$). Hence, another explanation is that the illumination projected on the sample is equivalent to the illumination associated with the same DMD pattern centered on the optical axis up to a phase ramp (as a consequence of Fourier transform properties). This defines a numerical aperture $N\!A$, if this aperture is low then the resolution of the collected images can only be low (see Figure~\ref{fig:ConfigON}).

\subsection{Difference with the acoustic experiment}

In the acoustic experiment depicted in~\cite{PRAppliedAcoustic_Noetinger}, the field originates from a point-like source at a finite distance, allowing it to be described as a spherical wave. In contrast, in the suggested optical experiment, the object resides in the focal plane of the objective, i.e. in the far-field. In the acoustic experiment, each loudspeaker emits successively so the modulation takes place at the level of the lens’s pupil whereas in the optical experiment the whole lens is contributing to the signal. Another important difference is the use of intensity detector.

\subsection{Observation of vortex-like images with intensity-only measurements}

\begin{figure}[t!] 
    \begin{center}
        \includegraphics[width=0.8\columnwidth]{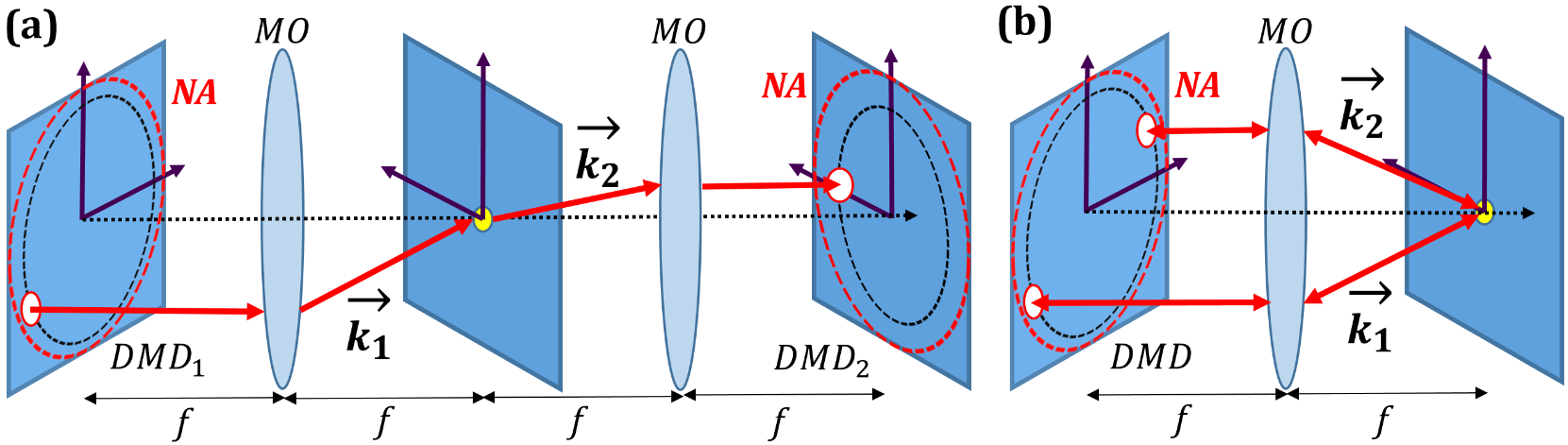}
        \caption{\label{fig:ConfigDMDtomo} \small{\textbf{Use of a symmetric sequence}}. \textbf{a)} The sample is in the Fourier plane of 2 different DMD: in the illumination path ($DMD_1$) and in the collection path ($DMD_2$). If the two DMD (depicted here in transmission) display a single point, the sample receives a plane wave directed by $\overrightarrow{k_1}$ and the transmitted field directed by $\overrightarrow{k_2}$ is collected. \textbf{b)} In reflection, displaying a sequence with two points allows to explore the same frequency space, if the two points are placed symmetrically relatively to the axis the field is real.}
    \end{center}
\end{figure}

To improve our understanding of the experiment proposed in this paper, let us consider the thought experiment depicted in Figure~\ref{fig:ConfigDMDtomo}.\textbf{a)} where the sample is illuminated by a plane wave of wavevector $\overrightarrow{k_1}$ using a DMD displaying a single rotating point in the Fourier plane of the sample and the transmitted field is collected only in the single direction of another plane wave $\overrightarrow{k_2}$ using a second DMD. This is rigorously the configuration of diffractive tomography~\cite{Tomoreview}. The frequency space explored is given by the sum of the two vectors $\overrightarrow{k_1}+\overrightarrow{k_2}$ and is limited by $\frac{2 N\!A}{\lambda}$ but the field is complex. To capture the same information in reflection, we propose to use a sequence with two rotating points. The PSF writes:
\begin{equation}
    \textrm{PSF}(\overrightarrow{r},t)=\left(e^{i\overrightarrow{k_1}(t)\cdot r}+e^{i\overrightarrow{k_2}(t)\cdot r}\right)\left(e^{-i\overrightarrow{k_1}(t)\cdot r}+e^{-i\overrightarrow{k_2}(t)\cdot r}\right)\propto 1+\cos\left((\overrightarrow{k_1}-\overrightarrow{k_2})\cdot \overrightarrow{r}\right)
\end{equation}
Thus, with this kind of pattern, it is possible to extract high spatial frequencies of the sample, here the term $\cos\left((\overrightarrow{k_1}-\overrightarrow{k_2})\cdot \overrightarrow{r}\right)$ but a low frequency term is also present (the term 1 here). Each wavevector being limited by $N\!A/\lambda$ the maximum frequency of the sum $2N\!A/\lambda$ is reached for a symmetric sequence of two points rotating symmetrically at the edge of the pupil similarly to the acoustic experiment~\cite{PRAppliedAcoustic_Noetinger}. In the case of the sector in Figure~\ref{fig:Rot45degPSF}.\textbf{a)}, for any pair of points of the pattern $|\overrightarrow{k_1}-\overrightarrow{k_2}|<N\!A/\lambda$ hence the decrease in resolution compared to the full-field configuration. In the case of two points rotating symmetrically, the illumination is then a rotating intensity fringe pattern as for structured illumination~\cite{StructuredIll}. In the general case, the 2D Fourier transform of a symmetric pattern being real, the PSFs are real and can be observed with an intensity-sensitive device. As a consequence of this symmetry, odd-numbered harmonics are lost.

\subsection{Preservation of optical sectioning: PSF with complementary pattern} \label{subs:Comp}

Illuminating the object by a rotating source and collecting the backscattered signal with a rotating receiver as in the acoustic setup can be achieved here by displaying a rotating point on the DMD. However, with this kind of sequence, the sample would be illuminated by a plane wave. This configuration is not favorable to imaging: a small amount of energy would be sent to a large area of the sample and an even smaller amount of energy would be collected and interfere. Moreover, optical sectioning, a key feature of a confocal microscope, would be lost.

\begin{figure}[b!] 
    \begin{center}
        \includegraphics[width=0.7\columnwidth]{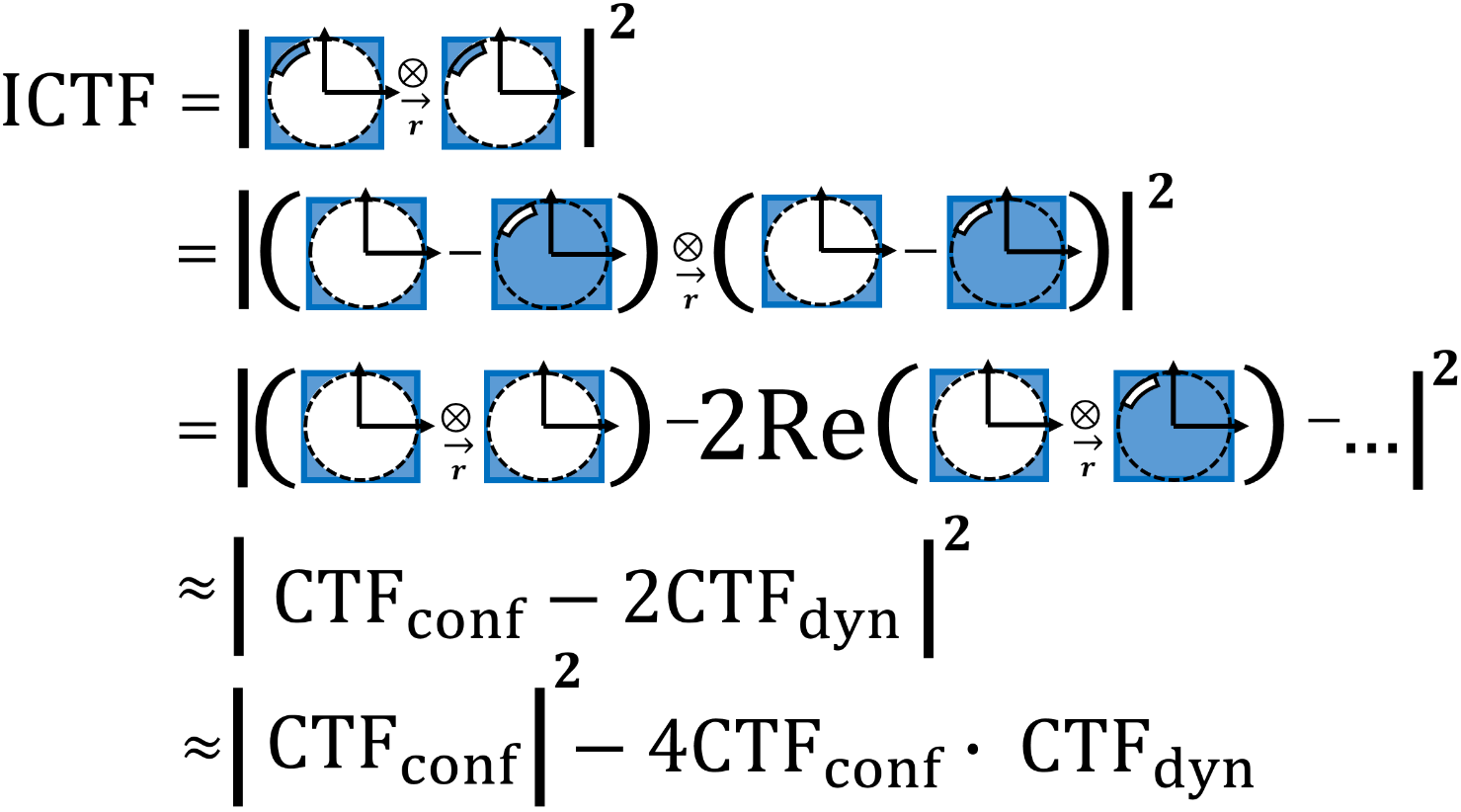}
        \caption{\label{fig:CTFcalc} \small{\textbf{ Associated intensity coherent transfer function (ICTF) to a complementary pattern.} Application of the formula $\textrm{CTF}=\textrm{CTF}_{\textrm{ill}}\otimes \textrm{CTF}_{\textrm{coll}}$, the complementary pattern can be decomposed as the substraction of the pattern of interest to the full pupil.}} 
    \end{center}
\end{figure}

In order to preserve optical sectioning its is better to focus the light on a specific spot. This means using the maximum surface of the pupil. We figured out that, instead of using a pattern associated to a time-dependent CTF named thereafter $\textrm{CTF}_{\textrm{dyn}}$, it is possible to use the complementary pattern which means the numerical aperture from which we substract the dynamic pattern. In order to understand the benefits of this particular type of patterns, the associated ICTF (the coherent transfer function \textit{in intensity}) associated to the measurements in the spatial Fourier space is illustrated schematically by the calculation of Figure~\ref{fig:CTFcalc}.

The new pattern can be expressed as the whole pupil minus the pattern associated to $\textrm{CTF}_{\textrm{dyn}}$. Applying the Equation 1 of the paper in Fourier space gives the new CTF. We assume that the product $\textrm{CTF}_{\textrm{dyn}} \otimes \textrm{CTF}_{\textrm{dyn}}$ can be neglected relatively to the other convolution products because the convolution with the full pupil amplifies the CTF. Then when the intensity is computed a term corresponding to the normal confocal microscope $|\textrm{CTF}_{\textrm{conf}}|^2$ shows up. This term is not time-dependent. The second term that shows up is proportional to the product $\textrm{CTF}_{\textrm{conf}}\cdot \textrm{CTF}_{\textrm{dyn}}$. The effect of $\textrm{CTF}_{\textrm{conf}}$ here is to amplify the signal that we seek which is $\textrm{CTF}_{\textrm{dyn}}$ more or less like an heterodyne amplification. The last term $|\textrm{CTF}_{\textrm{dyn}}|^2$ can be neglected. In addition we see in the calculation that we collect variations of the signal near $|\textrm{CTF}_{\textrm{conf}}|$ which can be positive or negative. This way we are sensitive to the field. 

Otherwise the ICTF would be $|\textrm{CTF}_{\textrm{dyn}}|^2$ which would be different than the signal we want to collect. For instance, the even harmonics would disappear. There is also a technical advantage with this configuration: the signal received by the photodiode is more intense compared to the situation with the original pattern where the flux would be close to zero. This means that the photodiode would be used near zero flux where the noise is more important than close to saturation as it can be used with a complementary pattern.  

As a consequence the image at the carrier frequency is more or less identical to the confocal image since the first term dominates at this frequency. Although with this type of pattern we benefit from the focusing, the drawback  is that the resulting CTF is sensitive to the decreasing gain of $\textrm{CTF}_{\textrm{conf}}$ for high spatial frequencies. However, the multiplexing effect highlighted in this article makes it possible to isolate those components with a sensitivity relatively high compared to the noise.

\section{Numerical simulation}

\subsection{Principle}

Using relation~(\ref{PSFconf}) in the Fourier space, the time-dependent CTF can be computed from the pattern. The illumination and collection PSFs correspond both to the 2D Fourier transform of the pattern $M$. From a stack of images to be displayed on the DMD, the temporal CTF can be computed:
\begin{equation}
    \textrm{CTF}(f_x,f_y,t)=M(f_x,f_y,t) \otimes M(f_x,f_y,t)\equiv M\left(\frac{2\pi x}{\lambda f},\frac{2\pi y}{\lambda f},t\right)\otimes M\left(\frac{2\pi x}{\lambda f},\frac{2\pi y}{\lambda f},t\right)\
\end{equation}
textcolor{red}{Tu es sûr des 2 $\pi$ ici? On n'est pas en train d'écrire en fréquence spatiale?}
Then the PSF can be computed with a 2D Fourier transform.

However it is not necessary to compute the PSF to get the predicted images, the CTF can be multiplied with the object in the Fourier domain to obtain $\widehat{Im}$. The inverse Fourier transform then gives the image $Im(x,y,t)$. The \textit{harmonic images} can be obtained using the temporal Fourier transform and then filter around $n\Omega$ or, more straightforwardly, by computing the coefficients of the Fourier series.

\subsection{Predicted images}

\begin{figure}[h!]
    \begin{center}
        \includegraphics[width=0.9\columnwidth]{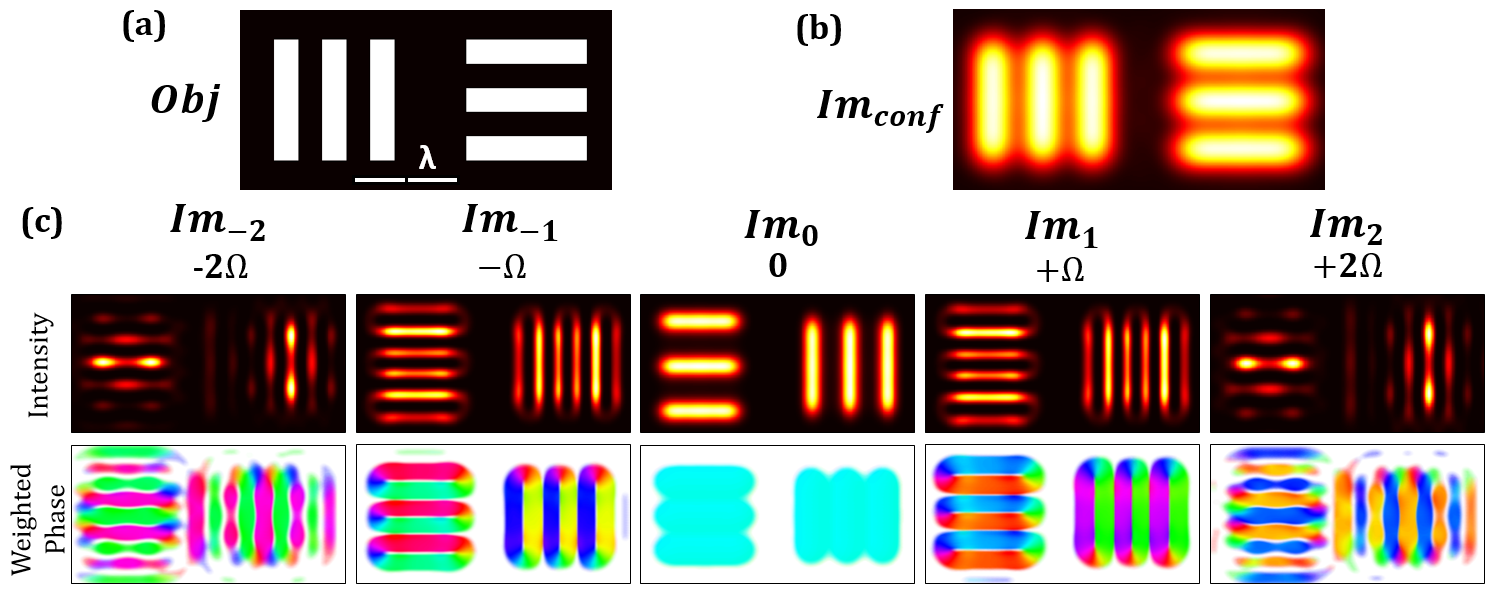}
        \caption{\label{fig:ImDynG11e1}\textbf{Predicted images} from the numerical simulation with 244nm lines. \textbf{a)} Object. \textbf{b)} Confocal image. \textbf{c)} Dynamic images $Im_n$ obtained with the full pupil deprived of a rotating sector at different harmonic frequencies $n \Omega$ for $n \in [\![-2;2]\!]$ with weighted phase and intensity.}
    \end{center}
\end{figure} 

On Figure \ref{fig:ImDynG11e1} we show an example of a numerically computed image of the 244~nm lines for comparison with the experimental data showed in the third figure of the main text.  
We see the qualitative agreement with the experimental images at $\pm\Omega$ as shown on the third figure of the main text. We also witness that the data at $\pm 2 \Omega$ from the simulation is qualitatively different from the experimental data.

\section{Experiment \& Results}

\subsection{Experimental setup}

\begin{figure}[b!]
    \begin{center}
        \includegraphics[width=1.02\columnwidth]{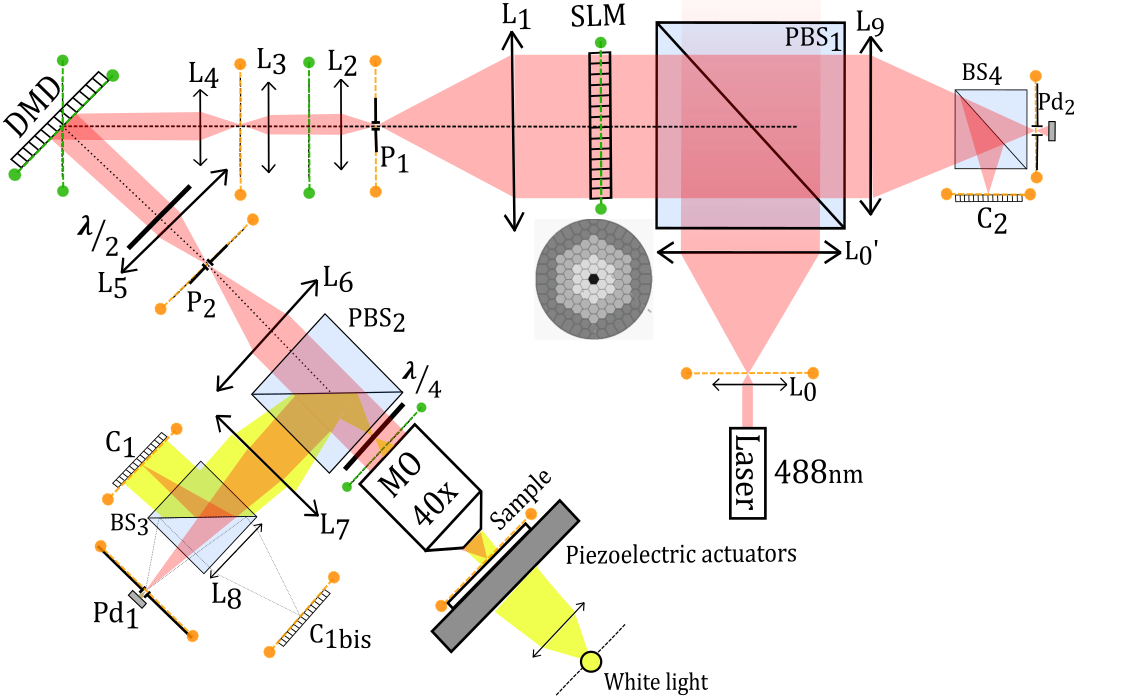}
        \caption{\label{fig:SchemaManipTot}\textbf{Final setup to produce dynamic confocal images.} The images obtained with each photodiode are named $|Im_1|^2$ or $|Im_2|^2$. \small{The focal planes of each lense are iidcate by dotted lines. Conjugated planes are indicated by orange or green dotted lines. 
        To go from one to the other using one lense 2D Fourier transform + scaling are performed.  Distances are not to scale. L: lenses, C:  CCD camera, PBS: polarizing beamsplitter, MO: Microscope objective, P: pinhole , Pd:photodiode behind a pinhole , BS:  beamsplitter, $\lambda/4$: quarter waveplate, $\lambda/2$: half waveplate, DMD: digital micromirror device, SLM:spatial light modulator, I: adjustable iris.}}
    \end{center}
\end{figure} 

The full experimental setup is depicted in Figure~\ref{fig:SchemaManipTot}. It is made using a polarized turquoise laser Coherent Sapphire SF NX @ 488~nm with a narrow-linewidth of 1.5~MHz (equivalent to 30~m of coherence length). The beam is enlarged with  beam expanders, filtered using a pinhole (P$_1$), sent to a \textit{Vialux} DMD with $1920 \times 1200$ pixels and then to the sample using an \textit{Olympus} MPLFLN40X microscope objective of numerical aperture $N\!A=0.75$ with a magnification adapted to use the full numerical aperture of the objective. The sample is also illuminated by white light (\textit{Kohler illumination}) in a conjugated plane of a camera (C$_1$) for sample positioning. By adjusting the light polarization with waveplates and polarizing beam splitters, the flux can be sent to one or the other photodiode (Pd$_1$ and Pd$_2$) to build dynamic confocal images. The \textit{Thorlabs} PDA10A2 photodiodes are placed behind pinholes whom equivalent size in the object plane is approximately 1 Airy unit. The current from the photodiodes is amplified by a tunable transimpedance amplifier and recorded by a  \textit{Picoscope} electronic oscilloscope. The backscattered field can also be observed with or without the DMD filtering using camera C$_1$ or C$_2$. To compensate for the aberrations induced by the surface of the DMD~\cite{DMDsetting} a \textit{Meadowlark} liquid-crystal spatial light modulator (SLM) with 127 hexagonal pixels is placed in a conjugated plane of the DMD. 
The DMD is more or less equivalent to a configurable blazed grating, this explains the two intersecting equiphase planes at the DMD level on Figure~\ref{fig:SchemaManipTot} of the main text. see~\cite{DMDsetting} for a detailed tutorial.\\

\subsection{Aberration correction}

\begin{figure}[t!]
    \begin{center}
        \includegraphics[width=0.7\columnwidth]{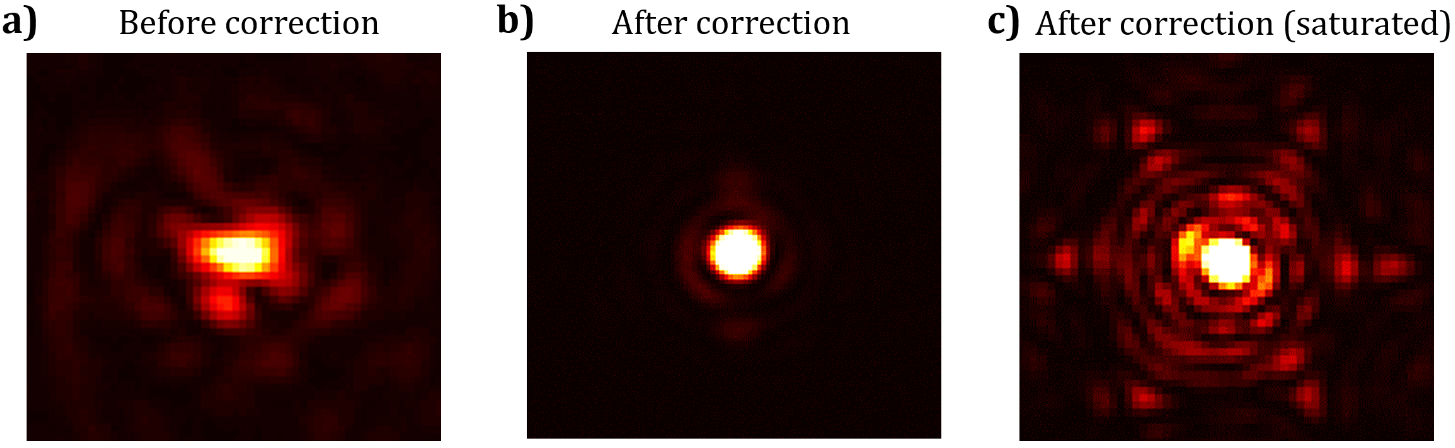}
        \caption{\label{fig:Correction}\textbf{Aberration correction} Images of the focal spot associated to a disk on the DMD (full numerical aperture) taken with camera $C_1$. Before and after correction.}
    \end{center}
\end{figure}

The aberrations come from the DMD surface and are smoothly varying. Hence, by reducing the surface of the DMD used it is possible to mitigate this effect. This is equivalent to a phase mask that makes the illumination \& collection PSFs different from the Airy spot.
See~\cite{DMDab} for a complete tutorial. In this tutorial, the pattern shown on the DMD is modified to retrieve a nice focal spot. Here, since we wish to use the DMD surface to project a time-varying pattern we preferred to add a SLM.\\
To compensate for this we place the \textit{Meadowlark} SLM in a conjugated plane. By varying the voltage and thus the phase difference of each pixel one after the other it is possible to maximize the intensity in a small zone and to retrieve a nice Airy spot as shown in Figure~\ref{fig:Correction}. The pixels of the SLM are hexagonal, this introduces some high frequencies in the Fourier plane of the sample. They remain visible in the sample's/camera plane when saturating the camera in the form of an hexagonal modulation (Figure \ref{fig:Correction}.\textbf{c}).

\subsection{Determination of the full-field and confocal resolution of the setup}

\begin{figure}[h!]
    \begin{center}
        \includegraphics[width=0.6\columnwidth]{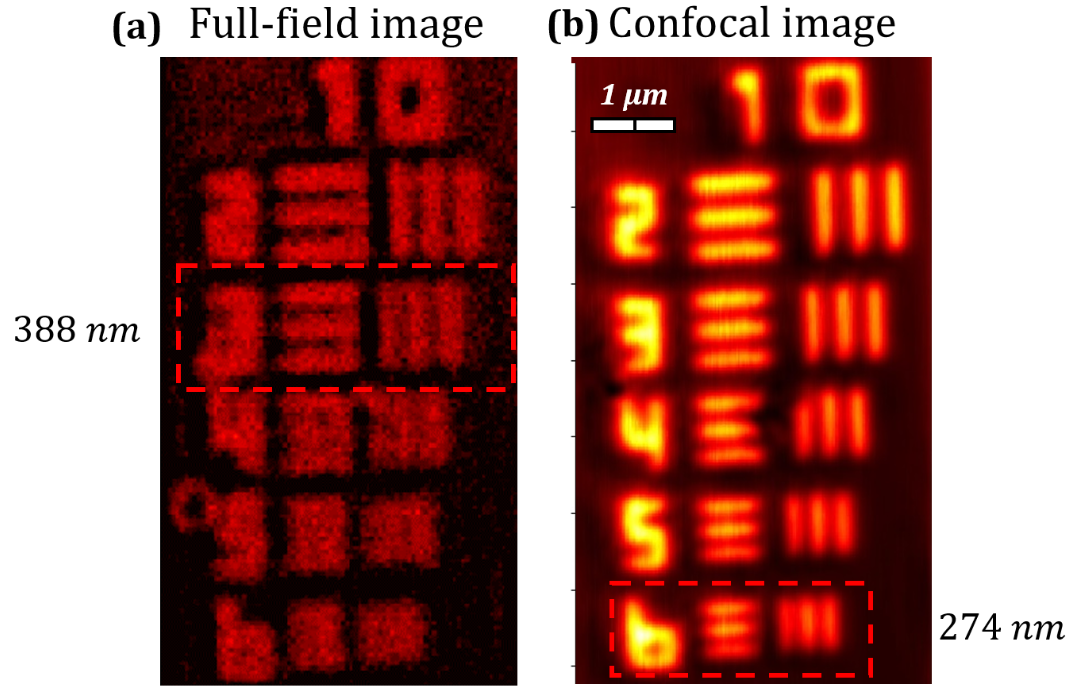}
        \caption{\label{fig:ResLimitUSAF}\textbf{Experimental resolution limit.} Full-field image acquired with camera C$_1$. Confocal image acquired with photodiode Pd$_2$ using the full numerical aperture.}
    \end{center}
\end{figure}

The resolution is determined roughly using experimental full-field and confocal images shown on Figure \ref{fig:ResLimitUSAF}.

\subsection{Choice of the rotating sequence}

\begin{figure}[t!]
    \begin{center}
        \includegraphics[width=0.8\columnwidth]{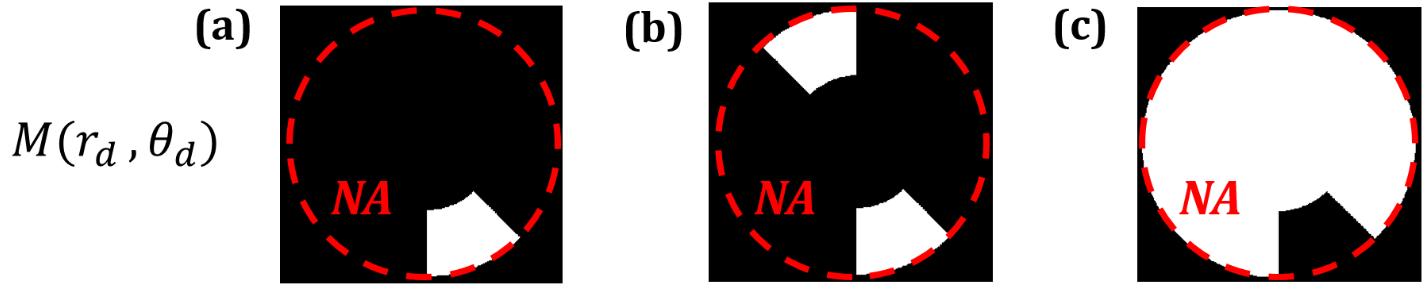}
        \caption{\label{fig:Motif}\textbf{Different kind of patterns to be displayed by the DMD} \textbf{a)} A sector of 45$^\circ$ truncated to collect only high spatial frequencies. \textbf{b)} The same duplicated pattern in order to do intensity only measurement. \textbf{c)} Complementary pattern to a), optical sectioning is preserved making the maximum use of the pupil, the same time-varying information is extracted as for a) and the signal is more powerful.  }
    \end{center}
\end{figure}

We consider different patterns to be displayed on the DMD as shown on Figure \ref{fig:Motif}.

The first one corresponds to the most intuitive choice for reproducing the acoustic experiment. As explained in~\ref{subs:Explations}, this choice is naive as it leads to a week illumination on a large area. Furthermore it would necessitate to record the field to build an image. The solution is to use interference either with a reference beam or with a beam coming from the sample. The second one allows to perform intensity measurement only as the 2D Fourier transform of a symmetrical centered function is real) and to capture high frequencies of the sample. The last one makes maximum use of the pupil and thus preserves the optical sectioning. In this case the PSFs of the system are more complex (see~\ref{subs:Comp}).

\begin{figure}[b!]
    \begin{center}
        \includegraphics[width=0.8\columnwidth]{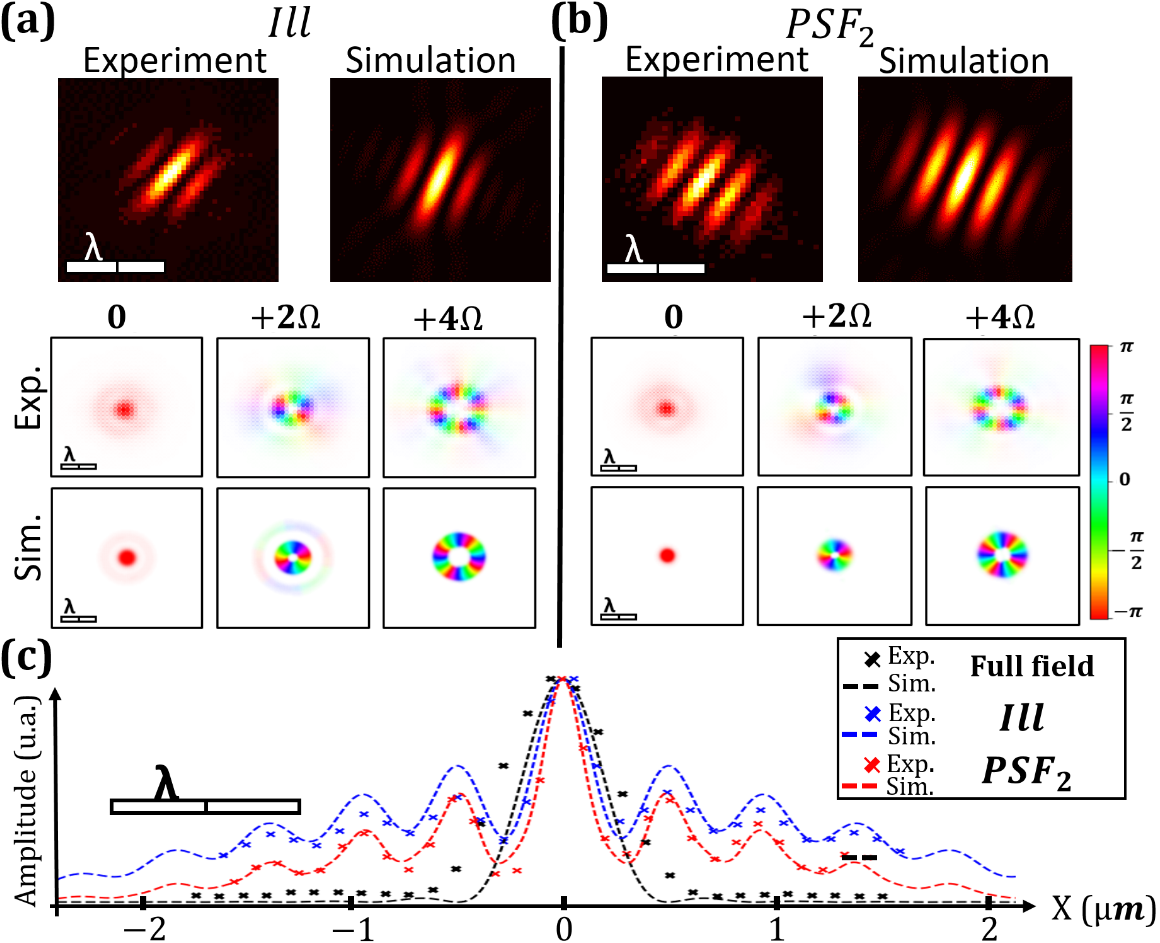}
        \caption{\label{fig:ObserVortexCam2} \textbf{Observation of the experimental vortices associated to a rotating sequence} \textbf{a)} Illumination of the sample $Ill$, comparison of the experiment (field on camera C$_1$) and numerical simulation with weighted phase.\textbf{b)} PSF$_2$, comparison of the experiment (field on camera C$_2$) and numerical simulation with weighted phase. \textbf{c)} Cross-section of the field at harmonic $0$, with a comparison of the experimental values (crosses) and simulation (dotted lines). }
    \end{center}
\end{figure}

\subsection{Observation of the experimental vortices}

To obtain an experimental proof that vortices can be obtained with a rotating sequence and to validate our model, we consider first a simple experiment. The sample is positioned in a homogeneous reflective area such as a glass plate.
Then, the field collected by the camera C$_1$ (resp. C$_2$) is the sample's illumination $Ill(r,\theta,t)$ (resp. $Ill(r,\theta,t)\cdot Ill(r,\theta+\pi,t)=\textrm{PSF}_2(r,\theta,t)$). The field on the first camera is the 2D Fourier transform of the DMD pattern whose reflection is collected by the objective in a full-field way. In a Fourier point of view, the pattern has necessarily a support limited by the numerical aperture, the full-field imaging operation is a multiplication by the aperture which has no effect here. The same explanation applies for the second camera with an additional multiplication by the DMD pattern.\\

In order not to saturate the camera and to observe an intensity pattern the pattern sent is a duplicated sector as depicted on Figure \ref{fig:Motif}.
As can be seen from Figure \ref{fig:ObserVortexCam2} there is a good agreement between the simulation and the experiment. For the signal recorded at $\omega_0$ by the second camera namely $\textrm{PSF}_{2,0}$.

\subsection{Experimental dynamic images}

\begin{figure}[h!]
    \begin{center}
        \includegraphics[width=0.6\columnwidth]{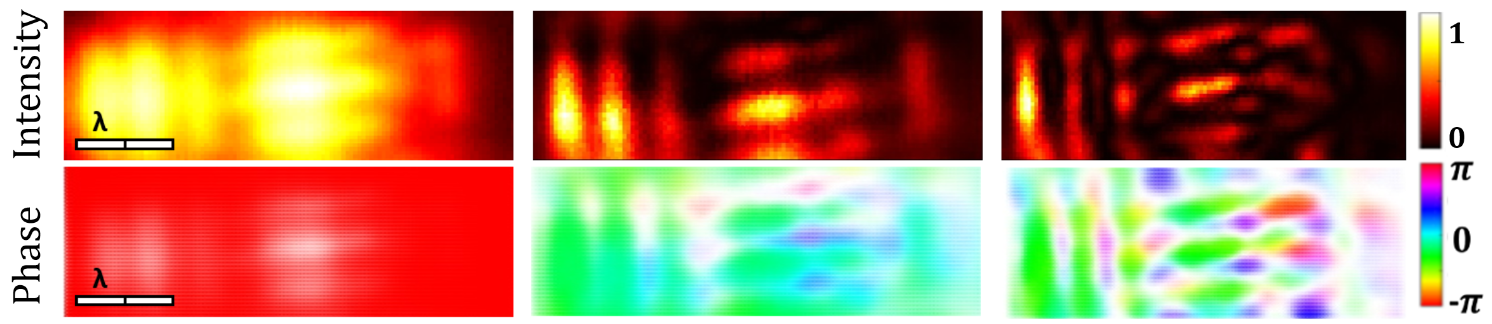}
        \caption{\label{fig:ImSect45dedoubleGr11e1}\textbf{Dynamic images of the first element of group 11 of the USAF target.} Linewidth is 244 nm }
    \end{center}
\end{figure}

With a doubled 45$^\circ$ sector as shown on Figure \ref{fig:Motif}.\textbf{b)} the images obtained do not correspond to the numerical simulation. An example of the typical data obtained is depicted on Figure \ref{fig:ImSect45dedoubleGr11e1}. We see that the intensity and phase variations follow the pattern of the target but without appearing like a convolution with a vortex-like function in opposition to what is observed on the images of the third figure of the main text using \textit{complementary sequences}. Our interpretation is that Equation \ref{PSFconf} only makes sense when there is \textit{optical sectioning} is present using relation \textit{ie.} when the light/signal collected comes form a narrow diffraction-limited spot.

\begin{figure}[h!]
    \begin{center}
        \includegraphics[width=0.6\columnwidth]{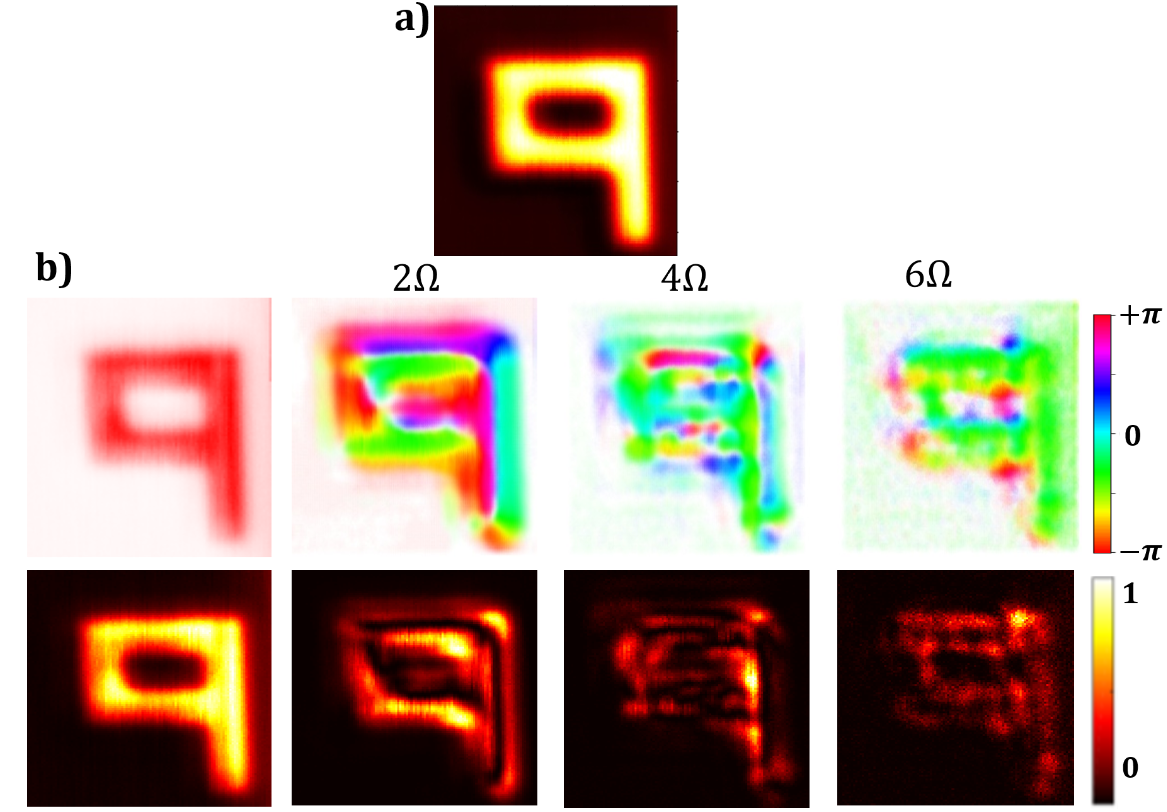}
        \caption{\label{fig:ImPD1}\textbf{Dynamic images of the first element of group 11 of the USAF target.} captured using the first photodiode Linewidth is 980 nm approximately. }
    \end{center}
\end{figure}

Although the images captured using the first photodiode where encouraging for large features no satisfying experimental data has been obtained for small features (Figure \ref{fig:ImPD1}). We have no explanation for this.

\section{Reconstruction of the images}

\subsection{Detailed procedure for \textit{dynamic} confocal images}

The experimental CTF is computed from the experimental sequence used using $\textrm{PSF}_{\textrm{conf}}=\textrm{PSF}_{\textrm{ill}} \cdot  \textrm{PSF}_{\textrm{coll}}$ and the classic relationship of Fourier optics  as:
\begin{equation}
    \textrm{CTF}_{2}(f_x,f_y,t)=M(f_x,f_y,t)\otimes M(f_x,f_y,t) \ \ .
\end{equation}
The pseudo-inverse is computed by applying the Tikhonov formula:
\begin{equation}
    \textrm{iCTF}_n=(\textrm{CTF}_n^* \cdot \textrm{CTF}_n + \sigma)^{-1}\cdot \textrm{CTF}_n^* \label{InvTikh}
\end{equation}
Then it is multiplied by a filter consisting of an annular aperture of width between $(1+f_{super})\frac{2N\!A}{\lambda}$ and $\frac{N\!A}{4\lambda}$ where $f_{super}$ is a tunable positive factor allowing for superresolution or not and canceling high frequency inversion artifacts. Using this filter we mitigate high frequency artifacts and the low frequency background. In all the results presented here $f_{super}$ was set between 0.5 to 1. Hence we use all the upper part of the optical spatial bandwidth.

For each acquisition, the data are apodized using a 2D Tukey window in order to reduce the high frequencies associated to the edges. After this operation, the 2D Fourier transform of each image $\widehat{Im_n}$ is computed. It is then multiplied by $\textrm{CTF}_n$ to obtain $Obj_n$. The reconstruction $Obj_{-1}$,$Obj_0$,$Obj_1$ are then summed.

\subsection{Choice of $\sigma$}

\begin{figure}[h!]
    \begin{center}
        \includegraphics[width=0.8\columnwidth]{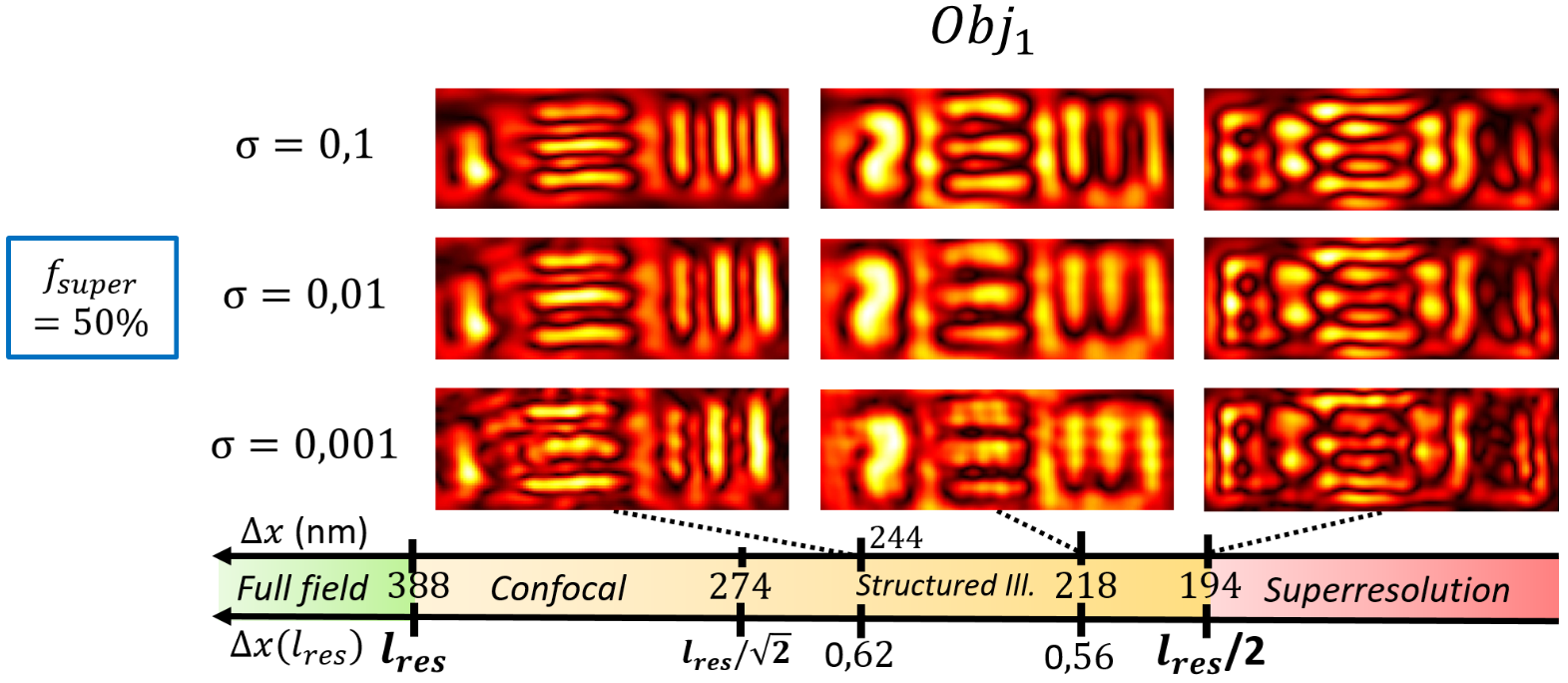}
        \caption{\label{fig:ChoixSigma}\textbf{Choice of $\sigma$.} By comparing different reconstructions (here the reconstruction associated to the image at $\Omega$ namely $Obj_1$) obtained with different values of $\sigma$ an appropriate value is chosen.}
    \end{center}
\end{figure}

The choice of $\sigma$ is performed by trying various values ranging from 0.1 to 0.001 and examining different reconstructions. The quality of the reconstruction is assessed qualitatively as shown on Figure \ref{fig:ChoixSigma}.

In order to obtain the best possible reconstruction, it is better to choose the highest value possible but to maintain  a reasonable level of artifacts. That is why we choose the value of $0.01$.

\subsection{Deconvolution of \textit{static} confocal images}

The confocal images acquired without any modulation of the pupil are deconvolved using the theoretical classical confocal PSF known as:
\begin{equation}
    \textrm{PSF}_{\textrm{conf}}(r) \propto \frac{J_1\left(\frac{2 \pi N\!A r}{\lambda}\right)}{\frac{2 \pi N\!A r}{\lambda}} \ \ .
\end{equation}
A pseudo-inverse is computed also using Tikhonov inversion (formula \ref{InvTikh}). Then, the deconvolution is also performed by multiplying this pseudo-inverse with $\widehat{Im_{\textrm{conf}}}$ after apodizing the image and filtering the pseudo inverse above the cut-off frequency $\frac{3 N\!A}{\lambda}$.

\end{document}